\documentclass[12pt,a4paper]{article}
\usepackage[utf8]{inputenc}
\usepackage[T1]{fontenc}
\usepackage{lmodern}
\usepackage[a4paper,margin=2.5cm]{geometry}
\usepackage{setspace}
\usepackage{microtype}
\usepackage{amsmath,amssymb}
\usepackage{booktabs}
\usepackage{threeparttable}
\usepackage{multirow}
\usepackage{array}
\usepackage{graphicx}
\usepackage{caption}
\usepackage{subcaption}
\usepackage{longtable}
\usepackage{float}
\usepackage[authoryear,round]{natbib}
\usepackage{xcolor}
\usepackage{authblk}

\usepackage[
    colorlinks=true,
    linkcolor=black,
    citecolor=black,
    urlcolor=blue
]{hyperref}

\title{\textbf{Corruption Perceptions and Revealed Behavior: Evidence from a Laboratory Experiment}}

\author[a]{Francesca Pancotto}
\author[b]{Enrico D'Ecclesiis}

\affil[a]{Dipartimento di Studi Linguistici e Culturali, Università degli Studi di Modena e Reggio Emilia (UNIMORE), Largo Sant'Eufemia 19, 41121 Modena, Italy}

\affil[b]{Dipartimento di Economia Marco Biagi (DEMB), Università degli Studi di Modena e Reggio Emilia (UNIMORE), Viale Berengario 51, 41121 Modena, Italy.

\textbf{Corresponding author}. Email: enricodecc@gmail.com}

\date{}

\begin{document}

\maketitle

\vspace{-0.5cm}

\begin{abstract}
Perception-based measures are widely used to assess corruption, yet the correspondence between general corruption perceptions and individuals' own behavior and experience may depend on the position from which corruption is encountered. We examine this relationship in a repeated bribery game embedded in a public-goods setting, with 396 participants interacting in fixed groups of three citizens and one judge. Participants subsequently report a survey-style assessment of how widespread corruption is in the country. Among citizens, perceived corruption shows no systematic relationship with their own bribery, contribution behavior, punishment experience, or the outcome of their own bribe offers; for bribery and punishment, associations as large as those estimated for judges can be excluded. Among judges, by contrast, bribes accepted and punishment inflicted are jointly associated with higher perceived corruption (joint test, $p=0.016$), and perceived corruption tends to increase gradually with the amount of bribery a judge accepts. A pooled test indicates that the correspondence between perceptions and own behavior is stronger for judges than for citizens (joint test, $p=0.029$; $p=0.018$), with stronger individual evidence for punishment than bribery. Judges who reject a larger share of bribe offers tend to report lower corruption, although this relationship is estimated less precisely. More broadly, the results suggest that the behavioral content of general corruption perceptions has a positional dimension: the same type of survey assessment may correspond differently to revealed behavior depending on respondents' exposure, participation, and decision authority.
\end{abstract}

\vspace{0.4cm}

\noindent\textbf{Keywords:} corruption; corruption perceptions; bribery; laboratory experiment; corruption measurement; information

\vspace{0.15cm}

\noindent\textbf{JEL Classification:} C91; C92; D73

\newpage

% ------------------------------------------------
% Main text
% ------------------------------------------------
\section{Introduction}
\label{sec:introduction}

Corruption is a central concern in political economy because it affects the allocation of public resources, the quality of government, economic performance, and citizens' relationship with public institutions \citep{Mauro1995,Svensson2005,RoseAckermanPalifka2016}. At the same time, corruption is inherently difficult to measure. Illegal transactions are deliberately concealed, administrative indicators can reflect enforcement as much as underlying corruption, and comparable objective measures are rarely available across institutional settings. Much empirical research consequently relies on perception-based indicators or survey assessments of corruption \citep{Kaufmann2005,Knack2007}. The headline international benchmarks are of this kind. Transparency International's Corruption Perceptions Index aggregates the perceptions of business people and country experts drawn from thirteen data sources, and the Control of Corruption indicator of the Worldwide Governance Indicators likewise combines perception-based sources into a single index \citep{KaufmannKraayMastruzzi2011}. It is useful to distinguish three kinds of measure. Declared perceptions, such as these indices, record what respondents believe about the prevalence of corruption. Declared experience, as collected for instance by Transparency International's Global Corruption Barometer or the World Bank Enterprise Surveys, records what respondents report having encountered, and remains subject to reticence and misreporting \citep{AzfarMurrell2009,Agerberg2022}. Direct measures observe corruption or its traces independently of what respondents say, for example through missing public funds, audits, physical measures of public works, or observed payments and rule violations \citep{ReinikkaSvensson2004,GoldenPicci2005,Olken2007,FismanMiguel2007,FerrazFinan2008,OlkenBarron2009}. Behavior in an incentivized experiment belongs to this third kind. This creates an important measurement question: to what extent do general corruption perceptions correspond to corruption that individuals actually encounter and to the behavior they themselves reveal when corrupt opportunities arise?

Experimental economics provides one way to observe corrupt behavior directly under controlled incentives. Since the bribery game introduced by \citet{Abbink2002}, laboratory studies have shown that decisions to offer, accept, and punish bribes respond systematically to economic incentives, institutional arrangements, and social context \citep{Cameron2009,BarrSerra2010}. An important concern is whether behavior observed in such settings captures meaningful variation beyond the laboratory. \citet{ArmantierBoly2013} address this issue by implementing closely related bribery experiments across laboratory and field environments in Burkina Faso and Canada and find substantial consistency in behavioral treatment effects across settings. More recently, \citet{ArmandCouttsVicenteVilela2023} combine a behavioral corruption game with independently observed evidence of real-world misappropriation among local leaders in Mozambique and show that experimental behavior contains information about corruption outside the game. Together, these studies support the use of incentivized experiments to capture meaningful variation in corrupt behavior.

A separate survey literature asks whether perceptions of corruption correspond to respondents' reported experience. Early micro-level evidence often points to a substantial gap. Using the 2004 Global Corruption Barometer, \citet{WeberAbramo2008} compare personal or household bribery experience with corruption assessments for more than 48,000 respondents across 60 countries. They find that bribery experience is generally a weak predictor of corruption perceptions, while different perception measures tend to be much more strongly related to one another. \citet{RoseMishler2010} reach a similar conclusion using the New Russia Barometer. Although corruption is perceived as widespread across Russian public services, the correlation between a generalized perception of corruption and reported bribery experience is only 0.09. Their multivariate analysis further shows that contact with public services is particularly important for explaining whether individuals pay bribes, whereas perceptions are more strongly associated with information obtained from friends, observations in the community, and the media. Experience and perception therefore need not simply be alternative measures of the same underlying phenomenon.

Cross-national evidence reinforces this point while also showing that the strength of the relationship varies considerably across contexts and measurement strategies. \citet{RazafindrakotoRoubaud2010} compare household reports of petty bureaucratic corruption from more than 35,000 respondents in eight African countries with assessments by approximately 350 experts and find systematic discrepancies, including substantial overestimation of corruption by experts and evidence of ideological bias in expert assessments. Using international microdata, \citet{DonchevUjhelyi2014} likewise show that direct corruption experience is only one of several determinants of perception-based corruption measures. Yet the evidence is not uniformly characterized by weak correspondence. Using individual-level evidence across European countries and regions, \citet{Charron2016} finds that citizens' reported experiences and perceptions of corruption correspond substantially more closely than much previous work had suggested, and that citizen assessments also display considerable agreement with expert measures. \citet{GutmannPadovanoVoigt2020}, using a broad cross-country microdata set, similarly find that first-hand bribery experience is robustly associated with perceived corruption, while also showing that individual characteristics and country-level conditions explain an important part of the remaining divergence. \citet{Mocan2008} constructs a country-level measure of the breadth of corruption from individuals' reports of whether officials asked them for bribes and finds that it is strongly related to widely used perception indices, although the relationship is nonlinear and some countries display substantial discrepancies. More recently, \citet{Crombach2026} shows in a panel of 45 democracies that electoral periods are associated with changes in household and expert corruption perceptions without corresponding changes in households' reported corruption experiences, further underscoring the distinction between perceived and experienced corruption. Overall, the evidence points to substantial heterogeneity in how closely corruption perceptions correspond to personal experience.

Other studies help explain why such divergence can arise. \citet{CorradoEtAl2023} use European microdata to construct distinct latent measures of perceived and experienced corruption and examine the factors associated with the gap between them. Their results show that information acquired beyond direct personal experience, including the broader information environment, contributes to the formation of corruption perceptions. Related work shows that corruption perceptions can be affected by political information, media exposure, and response processes independently of respondents' own experience \citep{Agerberg2022,PittalugaEtAl2024}. Moreover, even experience-based survey measures are not equivalent to directly observed behavior: questions about bribery concern illegal and socially sensitive conduct and may be subject to reticence or misreporting \citep{AzfarMurrell2009,Agerberg2022}. Thus, the survey literature provides rich evidence on perception--experience correspondence, but both sides of the comparison remain based largely on reported assessments or reported encounters.

Field evidence provides an alternative benchmark. \citet{Olken2009} compares villagers' corruption perceptions with an independently constructed measure of missing expenditures in Indonesian road projects. Perceptions do contain information about measured corruption, but the correspondence depends strongly on what villagers are able to observe, showing that visibility affects the informational content of corruption assessments. In a different field setting, \citet{OlkenBarron2009} directly observe more than 6,000 illegal payments made during trucking trips in Aceh, demonstrating both that corruption can be measured through behavior rather than perceptions and that opportunities and bargaining conditions systematically shape realized payments. Together, this evidence shows both that perceptions can contain genuine information about corruption and that the mapping between perceptions, exposure, and realized corrupt transactions is neither mechanical nor uniform.

A small number of studies come especially close to connecting survey information and experimentally observed corruption within the same individuals. \citet{CamposOrtiz2011} combines a laboratory bribery game with survey measures of participants' previous experience with bribery and their tolerance of corruption. Participants classified as having a more corruption-prone background make choices more conducive to bribery: citizens comply with fewer requirements and offer larger bribes, while inspectors with more corruption-prone backgrounds are more likely to request bribes. The study therefore provides important evidence that self-reported corruption experience and attitudes have behavioral content in an incentivized experiment. Its focus, however, is on previous bribery experience and normative tolerance rather than on whether a general assessment of the prevalence of corruption corresponds to the individual's own revealed behavior. More recently, \citet{BuonannoPucaSbriglia2026} combine a corruption experiment with post-experimental measures of perceived corruption in participants' municipalities of origin and external measures of local civic capital. They show that perceived corruption is systematically related to local civic-capital indicators and that civic capital and information about counterparts' geographic origin are associated with corrupt behavior in the experiment. Their analysis uses perceptions and experimental behavior to study the legacy of institutions and social-identity cues rather than the individual-level correspondence between a general corruption assessment and one's own corrupt choices. These studies bring survey information and experimental behavior into the same empirical setting, but do not make the correspondence between general corruption perceptions and individuals' own revealed behavior their primary object of analysis.

We examine this correspondence directly by linking the same participants' survey-style assessment of how widespread corruption is in the country to corruption-related behavior observed under incentives. Our experimental environment also allows us to ask whether this relationship differs with the position individuals occupy within the corrupt interaction. This possibility is suggested by experimental evidence showing that roles within corrupt exchanges can generate different behavioral responses. \citet{LevatiNardi2023}, for example, find that allowing third parties harmed by corruption to communicate reduces citizens' propensity to offer bribes but does not significantly alter officials' acceptance decisions. \citet{BuonannoPucaSbriglia2026} likewise find role-specific responses to information about counterparts' geographic origin, with officials' acceptance behavior responding differently from citizens' bribe offers. Related experimental work shows that institutional roles affect the attribution of responsibility for corrupt decisions \citep{MonteroPossajennikovVerbel2026}. These results suggest that corruption is not encountered from a homogeneous observational or decision-making position.

This distinction is central to our experiment. Citizens always observe and control their own contribution and bribery decisions and experience the judge's response; depending on the experimental condition, they may also receive information about other citizens' behavior. Judges, by contrast, systematically observe the contributions and bribe offers of all three citizens and determine whether each offered bribe is accepted or rejected and how punishment is allocated. They therefore combine broader exposure to others' behavior with greater decision authority over whether corrupt exchanges are completed. If general corruption assessments partly incorporate individuals' own behavior and experience, we should observe a systematic relationship between the two. This gives our first hypothesis: \textit{Hypothesis 1: General corruption perceptions are systematically associated with individuals' corruption-related behavior and direct experience.} The literature reviewed above also suggests that the informational content of experience depends on what individuals can observe, while experimental evidence indicates that institutional roles shape how actors participate in corrupt exchanges. We therefore expect systematic correspondence to emerge more clearly in the position that combines broader observation of the interaction with direct authority over corrupt transactions. This gives our second hypothesis: \textit{Hypothesis 2: Systematic correspondence between general corruption perceptions and revealed behavior is expected to emerge more clearly for individuals whose position within the interaction provides broader exposure to others' behavior and greater decision authority.}

We test these hypotheses in a repeated bribery game embedded in a public-goods setting, with 396 participants interacting in fixed groups of three citizens and one judge. Roles are randomly assigned and remain fixed throughout the interaction. Citizens decide how much to contribute to a common project and whether to offer transfers to the judge. The judge observes the contributions and offers of all citizens, decides whether each positive transfer is accepted or rejected, and can impose costly punishment. After completing the experimental tasks, every participant answers the same survey-style question about how widespread corruption is in the country. The design therefore combines a common general perception measure with directly observed, role-specific behavior generated within the same underlying interaction.

The results differ across positions. Among citizens, general corruption perceptions show no systematic relationship with bribes offered, contributions to the common project, punishment received, or the outcome of their own bribe offers. Among judges, by contrast, accepted bribery and punishment inflicted are jointly associated with higher perceived corruption; the association is most robust for punishment, and perceived corruption increases gradually with the amount of bribery accepted. A pooled specification that compares the two roles directly indicates that the correspondence is stronger for judges than for citizens. The amount of bribery offered to a judge is not by itself associated with the judge's perception, whereas judges who reject more of the offers they receive tend to report lower corruption; this evidence is less precise and reflects mainly the difference between judges who never reject an offer and those who do. Behavioral correspondence therefore emerges in the judge position but not in the citizen position.

These findings contribute to the measurement of corruption in two ways. First, they extend the literature comparing perceptions with reported or independently measured corruption by examining correspondence with behavior that is directly observed under incentives. In doing so, the paper complements rather than replaces survey and field approaches: the laboratory provides a controlled behavioral benchmark for asking what individual-level content a general corruption perception may contain. Second, the results introduce a positional dimension into this comparison. The behavioral content of a general corruption assessment is not uniform across actors participating in the same corrupt environment. The pattern differs across positions characterized by different observational scope, forms of participation, and decision authority. This suggests that interpreting perception-based corruption measures requires attention not only to whether respondents have encountered corruption, but also to the position from which those encounters are accumulated.

\section{Experimental Design and Measures}
\label{sec:design}

The experiment is described in full in \citet{PancottoFedericoRighiHauert2026}, who study the effects of the experimental conditions on behavior in the bribery game. This section presents the features of the design that matter for the present analysis: what participants in each role decide and observe, how the experimental conditions differ, and how perceived corruption and behavior are measured.

\subsection{Experimental Environment and Roles}
\label{subsec:environment}

The core experimental environment is a repeated bribery game embedded in a public-goods setting and based on the institutional structure of \citet{BuffatSenn2018}. Participants are anonymously matched into fixed groups of four, consisting of three Participants A, whom we refer to as \textit{citizens}, and one Participant B, whom we refer to as the \textit{judge}. Roles are randomly assigned at the beginning of the game and remain unchanged throughout the interaction. Depending on the session, the game lasts between 12 and 15 periods. Each period consists of three phases and combines a public-good decision with opportunities for bribery and punishment.

In the first phase, each participant receives an endowment of 20 experimental points. Each citizen decides how many points to contribute to a common project, with the remainder retained in a private account. The sum of citizens' contributions is multiplied by 1.6 and divided equally among the four group members, including the judge. Thus, while contributing increases total group earnings, each citizen faces an individual monetary incentive to retain points in the private account. The judge does not make a decision in this phase but observes the citizens' contributions before making the decisions assigned to the judge later in the period.

In the second phase, each citizen receives an additional endowment of 5 points and decides whether and how many of these points, from zero to five, to transfer to the judge. Each point transferred costs the citizen one point and, if subsequently accepted, increases the judge's earnings by one point. These transfers constitute the bribery opportunity in the experiment. The judge takes no action during this phase.

In the third phase, the judge observes separately each citizen's contribution to the common project and the transfer offered by that citizen. For each offer, the judge decides whether to accept or reject the transfer. An accepted transfer is credited to the judge, whereas a rejected transfer is returned to the citizen. The judge also receives 15 deduction points that can be allocated across citizens. Each deduction point costs the judge one point and reduces the targeted citizen's earnings by three points, with at most 10 deduction points assignable to any single citizen. The order in which citizens are displayed to the judge is randomized across periods, so that displayed positions cannot be used to track a particular citizen over time.

The experimental structure therefore places citizens and judges in systematically different positions within the same interaction. Citizens observe and control their own contribution and bribery decisions and subsequently experience the judge's response. The judge observes the contribution and bribery offer of every citizen and decides whether each offered transfer is accepted and how punishment is allocated. These differences in observational scope and decision authority are central to the comparison developed below. The experimental conditions modify the information and outside options available around this common structure, as described next. The instructions for the bribery-game task are summarized in the Online Appendix, Section N; the complete instructions, in the original Italian and in English translation, are reported in the Supplementary Material of \citet{PancottoFedericoRighiHauert2026}.

\subsection{Experimental Conditions}
\label{subsec:treatments}

The experiment comprises four between-session conditions that retain the common bribery-game structure described above while varying the information available to citizens and, in one condition, their possibility of withdrawing from the interaction in a given period. We refer to these conditions as Full Information, Control, Transparency, and Exit.

In the \textit{Full Information} condition, citizens receive detailed feedback about behavior observed within their group. At the end of each period, they are informed about the contributions of the other citizens to the common project, the deduction points assigned by the judge, and whether transfers offered to the judge were accepted or rejected. Citizens therefore observe corrupt transactions involving other members of the group in addition to their own experience.

The \textit{Control} condition restricts this information. Citizens observe their own outcomes together with the contributions of the other citizens and the deduction points assigned by the judge, but they are not informed about the transfers offered by the other citizens or whether those transfers were accepted or rejected. They can therefore observe cooperation and punishment within the group without directly observing bribery involving the other citizens.

The \textit{Transparency} condition introduces endogenous access to information. Beginning in the second period, before making their Phase 1 decision, citizens can pay three points to obtain information about the judge's decisions concerning the other citizens in the previous period. The information includes their contributions, transfers to the judge, whether those transfers were accepted or rejected, and the deduction points assigned by the judge. Citizens therefore decide whether to incur a monetary cost to obtain a broader view of previous interactions within the group.

Finally, the \textit{Exit} condition retains the endogenous-information mechanism of the Transparency condition and additionally allows citizens to withdraw from the interaction in a given period. A citizen who chooses the outside option receives the initial 20-point endowment and the additional 5 points, net of any information cost, and does not participate in the common-project, transfer, or punishment stages for that period. If one citizen withdraws, the common project is shared among the two remaining citizens and the judge, and the judge's available deduction points are reduced from 15 to 10. If two or more citizens withdraw, the common project is not implemented and transfers do not take place.

The four conditions therefore generate different informational environments around the same underlying corruption interaction. Citizens differ in the extent to which they can observe bribery involving other group members, while the judge's informational position remains broad across conditions because the judge observes the contributions and bribe offers of all participating citizens and makes the corresponding acceptance, rejection, and punishment decisions. These design differences are accounted for in the empirical analysis when examining the relationship between corruption perceptions and individual behavior.

\subsection{Corruption-Perception Measure}
\label{subsec:perception_measure}

The dependent variable in all main analyses is participants' reported perception of corruption, collected in the final questionnaire after completion of the experimental tasks. Participants were asked: ``How widespread do you think corruption is in this country?'' Responses were recorded on a four-category scale: (1) hardly any of the public offices are corrupt, (2) few public officials are corrupt, (3) most public officials are corrupt, and (4) almost all public officials are corrupt. We code responses from 1 to 4, so that higher values indicate a greater perceived prevalence of corruption. In our sample, observed responses range from 2 to 4.

We interpret this item as a survey-style general assessment of corruption rather than as a measure of corruption within the experimental group. This distinction is central to the analysis: the outcome concerns corruption in the country, whereas the behavioral measures capture participants' directly observed choices and experience within the experiment.

Because all participants answer the same general question after occupying different positions within the same corruption interaction, we can examine whether the behavioral correspondence of that assessment differs across those positions. Additional details on the construction and coding of the perception measure are reported in the Online Appendix, Section A.

\subsection{Behavioral Measures and Individual Characteristics}
\label{subsec:behavioral_measures}

The behavioral measures differ naturally across the two institutional roles. For citizens, we focus on three period-level measures capturing distinct dimensions of behavior and direct experience within the game. \textit{Bribe offered} is the number of points transferred to the judge in the bribery stage and ranges from 0 to 5. It provides the most direct measure of the citizen's own participation in the corrupt exchange. \textit{Contribution} is the number of points allocated to the common project, ranging from 0 to 20, and captures cooperative behavior within the institutional environment in which bribery takes place. \textit{Punishment received} measures the effective monetary punishment imposed on the citizen by the judge; observations corresponding to the special non-participation code used in the Exit condition are treated as missing. Taken together, these measures distinguish the citizen's direct participation in bribery, cooperative behavior toward the common project, and direct experience of punishment by the judge. Including contribution therefore allows us to assess whether any correspondence with general corruption perceptions is specific to corruption-related behavior or extends to cooperative behavior within the same institutional environment.

For judges, the corresponding period-level measures reflect the decisions associated with their role in the corrupt exchange. \textit{Bribes accepted} records the total amount of bribery accepted by the judge in each period and ranges from 0 to 15. \textit{Punishment inflicted} records the total number of deduction points allocated by the judge to citizens in that period. These measures capture, respectively, participation in completed bribery transactions and the exercise of punishment within the group. We use punishment inflicted as a measure of the judge's behavior in the game and do not analyze its determinants here; see \citet{PancottoFedericoRighiHauert2026}. Total bribes offered to the judge is not included among these main behavioral measures because it reflects bribery opportunities generated by citizens rather than a decision made by the judge. We introduce it separately in Section~\ref{subsec:judge_mechanisms} to distinguish exposure to bribery opportunities from the judge's response to those opportunities.

We complement these period-level measures with a judge-level indicator of revealed resistance to bribery. For each judge, the \textit{bribe rejection share} is defined as the proportion of positive bribe offers received that the judge rejects. Zero transfers are excluded because they provide no opportunity for the judge to reveal whether an offered bribe would be accepted or refused. A higher rejection share therefore captures a greater revealed propensity to resist available bribery opportunities. As robustness checks, we also consider the total number of positive bribes rejected and an indicator for whether the judge rejects at least one positive bribe offer; these alternative measures are reported in the Online Appendix, Section E.

The final questionnaire additionally records socio-demographic characteristics, indicators of civic and prosocial engagement, generalized trust, and the stated justifiability of several rule-breaking behaviors, including bribe acceptance, tax evasion, fare evasion, and claiming undeserved public benefits. The main controlled specifications include age and gender, together with association membership and voting status as indicators of civic engagement and blood-donation status as a broader prosocial indicator. Gender is self-reported in the final questionnaire; the response categories observed in the data are Female, Male, Not identified, and Not answering, and all of them are retained in the analysis. These controls allow us to assess whether the behavioral relationships persist after accounting for observable differences in demographic characteristics and civic or prosocial orientation. Generalized trust and the stated justifiability measures are added in robustness specifications to examine whether the main relationships also persist after accounting for a broader set of individual attitudes.

Additional details on the construction and coding of the corruption-perception outcome, behavioral measures, and individual characteristics used in the analysis are reported in the Online Appendix, Section A.

\subsection{Procedures and Sample}
\label{subsec:procedures}

The experiment was conducted at the VERA Laboratory for Experiments in Social Sciences (VERALabEx) at Ca' Foscari University of Venice between September and December 2023. The experimental application was programmed in Python using oTree \citep{Chen2016}. Participants were recruited from the student population at Ca' Foscari University of Venice through the Online Recruitment System for Economic Experiments (ORSEE) \citep{Greiner2015}. In accordance with the VERALabEx Regulations and Guidelines, the experimental application, including the description of the experiment and the procedure, was submitted to the Steering Committee. The Committee did not identify any critical aspects concerning ethics or privacy and therefore did not require further review by an Ethics Committee. The Executive Board of the Department of Economics at Ca' Foscari University of Venice reviewed the application in its meeting of 5 September 2023, approved partial financing of the experiment in line with the applicant's request, and endorsed the project. The decision of the Executive Board was subsequently ratified by the Department Council of the Department of Economics at Ca' Foscari University of Venice on 14 September 2023 (decision 299/2023).

The experiment followed a between-session design comprising 14 sessions. Sessions were randomized across experimental conditions, and the number of periods in each session was also randomized, ranging from 12 to 15, in order to limit end-of-experiment effects. Within each session, participants were anonymously matched into fixed groups of four for the bribery game, with roles randomly assigned at the beginning of the interaction and remaining fixed throughout all periods. The experimental design and the hypotheses on the effects of the experimental conditions were preregistered on the Open Science Framework \citep{FedericoPancottoRighi2023}. The analysis reported in this paper was not part of that preregistration: the two hypotheses stated in the Introduction and the specifications used to examine them were not preregistered. The experiment was designed to study how information and the possibility of exit affect bribery and cooperation. The effects of the experimental conditions on bribery, cooperation, and punishment are analyzed in \citet{PancottoFedericoRighiHauert2026}. The present paper uses data from the same experimental sessions and addresses a different question: the relationship between behavior in the bribery game and the corruption-perception item of the final questionnaire, which is not analyzed there.

Upon arrival, participants were registered and assigned to individual cubicles. Before the experiment began, all participants received information about the study and privacy and provided informed consent. Visual and verbal communication among participants was not permitted during the sessions. Instructions were displayed on participants' screens and read aloud before the experimental tasks, and written instructions remained accessible throughout the experiment. Before each incentivized task, participants completed comprehension questions and could proceed only after providing the correct answers.

The bribery game analyzed in this paper was the first task completed during a broader experimental session. It was followed by a Distribution Game, a Cognitive Reflection Test, the Interpersonal Reactivity Index, and a final questionnaire collecting socio-demographic characteristics, civic and attitudinal measures, and the corruption-perception outcome used in the present analysis. Participants received no feedback on the outcomes of the other tasks before completing the questionnaire. Each participant took part in only one experimental session. Sessions lasted approximately 90 minutes. Experimental earnings were paid in cash, with average payments of approximately 15 euros for citizens and 18 euros for judges.

The sample used in this paper consists of 396 participants across the 14 experimental sessions, corresponding to 99 fixed bribery-game groups of three citizens and one judge. Two sessions were conducted under the Full Information condition, four under Control, four under Transparency, and four under Exit. This yields 36 participants in Full Information, 120 in Control, 124 in Transparency, and 116 in Exit. Since roles remain fixed within the bribery game, the sample contains 297 citizen roles and 99 judge roles. Two citizens in the Exit condition choose the outside option in every period, so that behavior is observed for 295 citizens and for all 99 judges. Detailed descriptive statistics and a summary of the instructions for the bribery-game task are reported in the Online Appendix, Sections B and N, respectively.

\section{Empirical Strategy}
\label{sec:strategy}

The analysis examines the two hypotheses developed in the Introduction. For Hypothesis 1, we estimate whether final corruption perceptions are systematically associated with role-specific behavior and experience. For Hypothesis 2, we compare these associations across citizens and judges, whose positions differ in observational scope and decision authority. Corruption perception is measured once for each participant after the experimental tasks, whereas the behavioral measures are generated repeatedly across periods. The unit of analysis is therefore the participant: each period-level behavioral measure is averaged within participant over all periods of the bribery game in which the participant takes part in the interaction. Periods in which a citizen chooses the outside option, and periods of the Exit condition in which the common project is not implemented because two or more citizens withdraw, involve no contribution, bribery, or punishment decisions and are treated as missing for both roles. The bribe rejection share is constructed over the same periods and varies only across judges. Earlier analyses of these data excluded the first period, in which the information-purchase option of the Transparency and Exit conditions is not yet available. Since that option is not the object of this paper, behavior is measured over all periods, and results that exclude the first period are reported in the Online Appendix, Section L.

For citizens, the main specification is
\begin{equation}
Corruption_i = \alpha + \boldsymbol{\beta}'\overline{Behavior}_{i} + \boldsymbol{\delta}'Condition_s + \varepsilon_{i},
\end{equation}
where $Corruption_i$ denotes the final corruption perception reported by participant $i$, $\overline{Behavior}_{i}$ contains the participant's average bribe offered, contribution to the common project, and punishment received, and $Condition_s$ is a set of experimental-condition indicators for session $s$. For judges, we estimate the analogous specification using average bribes accepted and punishment inflicted. The main analysis includes the role-specific behavioral measures jointly, while separate specifications for each measure are reported in the Online Appendix, Section D. For the joint specifications, we also report Wald tests of the null hypothesis that the behavioral coefficients are jointly equal to zero.

To evaluate Hypothesis 2, we estimate a pooled specification in which behavior and all controls are interacted with an indicator for the judge role, so that the role-specific coefficients coincide with those of the separate regressions and the interaction terms measure the difference between roles. Because the behavioral measures are role-specific and expressed in different units, bribery (offered for citizens, accepted for judges) and punishment (received for citizens, inflicted for judges) are standardized within role; the interaction coefficients therefore measure the difference between judges and citizens in the change in perceived corruption associated with a one-standard-deviation change in behavior. We report the two differences and a Wald test of the null hypothesis that both are zero. The way in which behavior is compared across roles was not specified in advance. We use standardized units for the main comparison because the original units differ in range and meaning across roles; the same comparison in original units, which provides weaker evidence of a difference, and a comparison based on a group-level measure common to the four group members are reported in the Online Appendix, Section K.

All main specifications include indicators for the experimental condition, which account parsimoniously for differences in the informational and institutional environments generated by the four conditions. For each analysis, we report a baseline specification with experimental-condition indicators and a second specification that adds age, gender, association membership, voting status, and blood-donation status. Because differences across experimental conditions and the coefficients on background characteristics are not the focus of the analysis, these coefficients are suppressed in the main tables and reported in full in the Online Appendix.

Specifications are estimated by OLS with heteroskedasticity-robust standard errors. In the pooled specification standard errors are clustered at the fixed-group level, because the citizens and the judge of a group share the same history of interaction. We assess robustness by retaining the period-level observations with participant-clustered standard errors, accounting for the ordinal nature of the outcome, adding broader attitudinal controls, clustering at the fixed-group level, replacing experimental-condition indicators with session fixed effects, and varying the definition of the estimation sample, including the exclusion of the first period. For judges, we additionally separate bribery opportunities received from bribery actually accepted and examine alternative measures of resistance to bribery. For citizens, we additionally examine the outcome of their own bribe offers. The corresponding results are reported in the Online Appendix.

\section{Results}
\label{sec:results}

\subsection{Corruption Perceptions and Revealed Behavior across Roles}
\label{subsec:across_roles}

We begin by asking whether general corruption perceptions are associated with the behavioral measures observed in each role. Table \ref{tab:behavior_across_roles} reports the main specifications for citizens and judges, using one observation per participant, together with the difference between the two roles. Column (1) includes the role-specific behavioral measures together with experimental-condition indicators, while column (2) adds background characteristics. Complete estimates and the separate specifications for each behavioral measure are reported in the Online Appendix, Sections C and D.

{\small
\begin{longtable}{@{}lcc@{}}
\caption{Perceived corruption and revealed behavior across roles}
\label{tab:behavior_across_roles}\\
\toprule
& (1) & (2) \\
& Baseline & + Background \\
\midrule
\endfirsthead

\multicolumn{3}{l}{\small\textit{Table \thetable\ continued from previous page}}\\
\toprule
& (1) & (2) \\
& Baseline & + Background \\
\midrule
\endhead

\midrule
\multicolumn{3}{r}{\small\textit{Continued on next page}}\\
\endfoot

\bottomrule
\endlastfoot

\multicolumn{3}{l}{\textit{Panel A. Citizens}} \\
\addlinespace

Bribe offered
& -0.0101
& 0.0005 \\
& (0.0292)
& (0.0276) \\

Contribution
& -0.0027
& -0.0034 \\
& (0.0071)
& (0.0067) \\

Punishment received
& 0.0022
& -0.0002 \\
& (0.0114)
& (0.0102) \\

\addlinespace
Joint test of behavioral coefficients, $p$-value
& 0.907
& 0.962 \\

Participants
& 295
& 295 \\

\midrule

\multicolumn{3}{l}{\textit{Panel B. Judges}} \\
\addlinespace

Bribes accepted
& 0.0412$^{*}$
& 0.0503$^{**}$ \\
& (0.0239)
& (0.0235) \\

Punishment inflicted
& 0.0687$^{**}$
& 0.0728$^{**}$ \\
& (0.0273)
& (0.0297) \\

\addlinespace
Joint test of behavioral coefficients, $p$-value
& 0.016
& 0.014 \\

Participants
& 99
& 99 \\

\midrule

\multicolumn{3}{l}{\textit{Panel C. Difference between roles (judges $-$ citizens), standardized behavior}} \\
\addlinespace

Bribery: accepted vs.\ offered
& 0.1318$^{*}$
& 0.1449$^{*}$ \\
& (0.0753)
& (0.0744) \\

Punishment: inflicted vs.\ received
& 0.1498$^{**}$
& 0.1660$^{**}$ \\
& (0.0705)
& (0.0701) \\

\addlinespace
Joint test of the two differences, $p$-value
& 0.029
& 0.018 \\

Participants
& 394
& 394 \\

Fixed-group clusters
& 99
& 99 \\

\midrule

Experimental-condition controls
& Yes
& Yes \\

Background controls
& No
& Yes \\

\end{longtable}
}

{\footnotesize\noindent\textit{Notes:} OLS estimates using one observation per participant. The dependent variable is perceived corruption, coded from 1 to 4, with higher values indicating greater perceived corruption. Behavioral measures are averaged within participant over all periods of the bribery game in which the participant takes part in the interaction. Periods of the Exit condition in which the common project is not implemented because two or more citizens withdraw are treated as missing. Panels A and B report separate regressions for each role, with all role-specific behavioral measures entered jointly and robust standard errors in parentheses. Panel C reports the difference between the judge and the citizen coefficients from a single pooled regression in which behavior and all controls are interacted with a judge indicator; in that regression bribery (offered for citizens, accepted for judges) and punishment (received for citizens, in deduction points; inflicted for judges) are standardized within role, so that coefficients measure the change in perceived corruption associated with a one-standard-deviation change in behavior, and standard errors are clustered at the fixed-group level. Column (1) includes experimental-condition indicators. Column (2) additionally controls for age, gender, association membership, voting status, and blood-donation status. Coefficients on controls are suppressed; complete specifications are reported in the Online Appendix, Sections C and K, and period-level specifications in Section G. The joint-test rows report $p$-values from Wald tests that the coefficients in the corresponding panel are jointly equal to zero. $^{*}p<0.10$, $^{**}p<0.05$, $^{***}p<0.01$.\par}

\hspace{1cm}

Panel A shows no systematic association between citizens' corruption perceptions and their behavior or direct experience in the experiment. The coefficients on bribes offered, contributions, and punishment received are all close to zero and remain so after background characteristics are added. The joint tests also fail to reject the null that these behavioral measures are unrelated to perceived corruption ($p=0.907$ and $p=0.962$). The citizen results therefore provide no evidence of the perception--behavior relationship anticipated in Hypothesis 1.

The results are different for judges. Punishment inflicted is positively associated with perceived corruption in both specifications ($p=0.013$ and $p=0.016$). The coefficient on bribes accepted is also positive; it is estimated less precisely in the baseline specification ($p=0.089$) and is statistically significant after background characteristics are included ($p=0.036$). The joint null is rejected in both specifications ($p=0.016$ and $p=0.014$). Hypothesis 1 therefore receives support in the judge position but not among citizens. Panel C compares the two roles directly. In standardized units, the association of perceived corruption with punishment is larger for judges than for citizens ($p=0.036$ and $p=0.020$), while the difference for bribery is positive but estimated less precisely ($p=0.083$ and $p=0.054$). The null hypothesis that both differences are zero is rejected ($p=0.029$ and $p=0.018$). The contrast between the two roles is consistent with Hypothesis 2, since the associations appear in the position with broader observation of others' behavior and direct authority over bribery and punishment decisions.

\begin{figure}[H]
\centering
\caption{Perceived corruption by level of own bribery}
\label{fig:perception_tercile}
\includegraphics[width=0.95\textwidth]{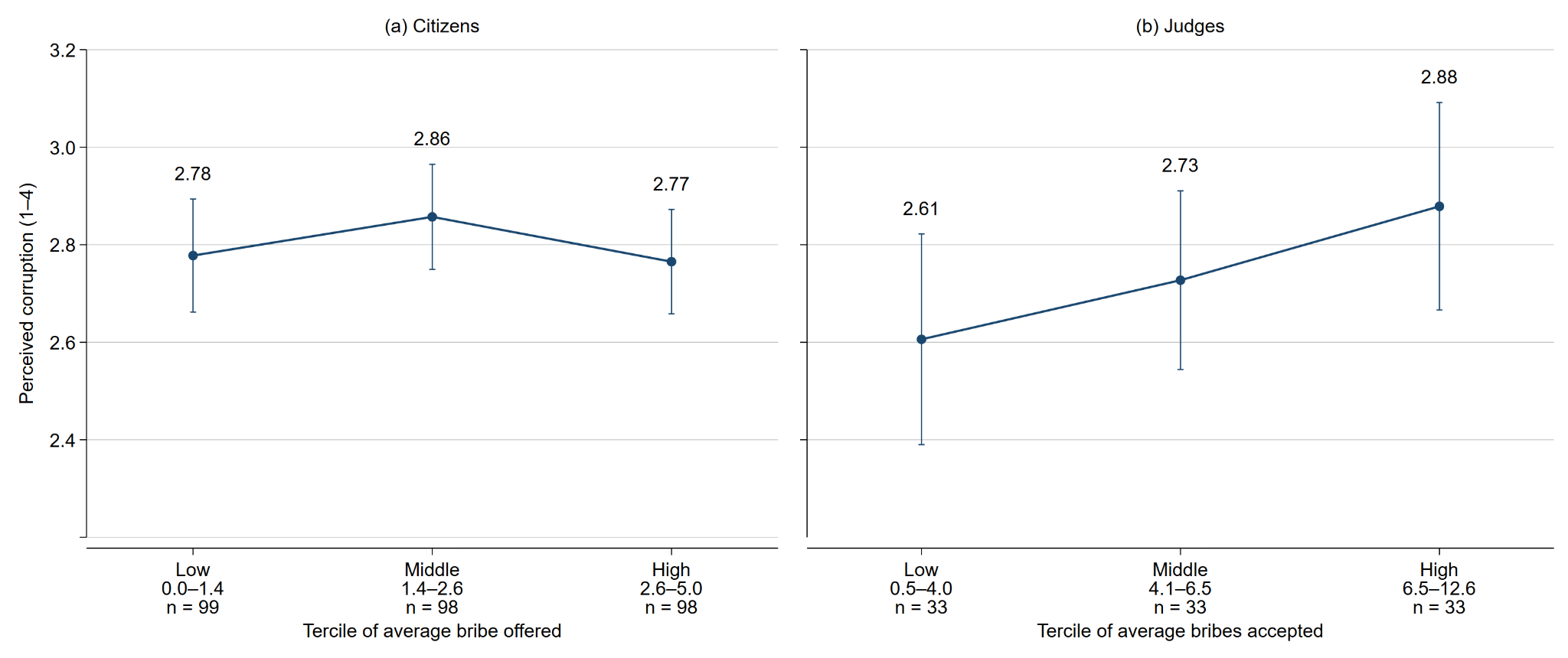}

\vspace{0.15cm}

\parbox{\textwidth}{\footnotesize\textit{Notes:} Average perceived corruption (coded from 1 to 4) with 95 percent confidence intervals, by tercile of the participant's average bribery over all periods of the bribery game in which the participant takes part in the interaction. Panel (a) groups the 295 citizens by the average bribe offered (0 to 5 points per period). Panel (b) groups the 99 judges by the average amount of bribes accepted (0 to 15 points per period). The range of the grouping variable and the number of participants in each group are reported below each label. The figure is descriptive and does not adjust for the experimental condition.}
\end{figure}

Figure~\ref{fig:perception_tercile} illustrates this pattern descriptively. Participants in each role are divided into three groups of equal size according to their average bribery in the experiment: the average bribe offered for citizens and the average amount of bribes accepted per period for judges. Among citizens, average perceived corruption is essentially the same across the three groups (2.78, 2.86, and 2.77). Among judges, it increases with the amount of bribery accepted, from 2.61 in the lowest third to 2.73 in the middle third and 2.88 in the highest third (Spearman rank correlation between group and perception: $p=0.055$). The share of judges answering that few public officials are corrupt falls from 45 percent in the lowest third to 24 percent in the highest third. In a regression of perceived corruption on indicators for the middle and highest thirds and experimental-condition indicators, judges in the highest third report 0.30 points more perceived corruption than judges in the lowest third ($p=0.055$), while the two indicators are not jointly significant ($p=0.156$). The increase is gradual rather than concentrated among the judges who accept the most: when a quadratic term in bribes accepted is added to the linear specification with experimental-condition indicators, it is not statistically significant ($p=0.239$). The group means and these tests are reported in the Online Appendix, Section B. The grouping into thirds is used for illustration only; the corresponding tests are those reported in Table~\ref{tab:behavior_across_roles}.

\subsection{Exposure and Active Responses in the Judge Position}
\label{subsec:judge_mechanisms}

The judge results allow us to distinguish more closely between exposure to bribery opportunities and judges' responses to them. Panel A adds the total amount of bribery offered to the judge alongside the amount accepted, while Panel B examines the decision to reject positive bribe offers. Table \ref{tab:judge_exposure_response} reports the results.

{\small
\begin{longtable}{@{}lcc@{}}
\caption{Judges' exposure and responses to bribery opportunities}
\label{tab:judge_exposure_response}\\
\toprule
& (1) & (2) \\
& Baseline & + Background \\
\midrule
\endfirsthead

\multicolumn{3}{l}{\small\textit{Table \thetable\ continued from previous page}}\\
\toprule
& (1) & (2) \\
& Baseline & + Background \\
\midrule
\endhead

\midrule
\multicolumn{3}{r}{\small\textit{Continued on next page}}\\
\endfoot

\bottomrule
\endlastfoot

\multicolumn{3}{l}{\textit{Panel A. Bribery opportunities and accepted bribes}} \\
\addlinespace

Bribes accepted
& 0.1082$^{***}$
& 0.1030$^{***}$ \\
& (0.0336)
& (0.0324) \\

Total bribes offered
& -0.0770$^{**}$
& -0.0609$^{*}$ \\
& (0.0349)
& (0.0360) \\

Punishment inflicted
& 0.0770$^{***}$
& 0.0788$^{***}$ \\
& (0.0251)
& (0.0263) \\

\midrule

\multicolumn{3}{l}{\textit{Panel B. Rejection of positive bribe offers}} \\
\addlinespace

Bribe rejection share
& -0.6156$^{*}$
& -0.5836$^{**}$ \\
& (0.3100)
& (0.2839) \\

\midrule

Experimental-condition controls
& Yes
& Yes \\

Background controls
& No
& Yes \\

Judges
& 99
& 99 \\

\end{longtable}
}

{\footnotesize\noindent\textit{Notes:} OLS estimates using one observation per judge, with robust standard errors in parentheses. The dependent variable is perceived corruption, coded from 1 to 4, with higher values indicating greater perceived corruption. Behavioral measures are averaged within participant over all periods of the bribery game in which the participant takes part in the interaction. Periods of the Exit condition in which the common project is not implemented because two or more citizens withdraw are treated as missing. Panel A jointly includes bribes accepted, total bribes offered to the judge, and punishment inflicted. Total bribes offered is the sum of the offers made by the citizens to the judge in a period. Because accepted bribes equal offered bribes minus rejected bribes, the coefficient on bribes accepted in Panel A is identical, with the opposite sign, to the coefficient on the amount rejected in a regression that includes the amount rejected and the amount offered. Panel B uses the bribe rejection share, defined as the proportion of positive bribe offers received over all periods of the bribery game that the judge rejects; zero offers are excluded because they provide no opportunity to accept or reject a bribe. Column (2) additionally controls for age, gender, association membership, voting status, and blood-donation status. Coefficients on controls are suppressed. Complete specifications are reported in the Online Appendix, Sections C and J, and alternative measures of bribe rejection in Section E. $^{*}p<0.10$, $^{**}p<0.05$, $^{***}p<0.01$.\par}

\hspace{1cm}

Panel A adds the total amount of bribery offered to the judge. Because the amount accepted equals the amount offered minus the amount rejected, the coefficients in Panel A have a direct reading: holding the amount offered constant, a judge who accepts one point more is a judge who rejects one point less. The coefficient on bribes accepted (0.1082 and 0.1030) is therefore the coefficient on the amount rejected with the opposite sign. Written in terms of the amount offered and the amount rejected, the same specification gives a coefficient of $-0.1082$ ($p=0.002$) on the amount rejected and of 0.0312 ($p=0.218$) on the amount offered; entered alone, the amount offered is likewise unrelated to perceived corruption ($p=0.354$). The amount of bribery that citizens offer to a judge is thus not significantly associated with the judge's perception, whereas the amount the judge turns down is. Two qualifications apply. First, the amount rejected is strongly right-skewed: the coefficient is roughly halved and is no longer statistically significant when the three judges who reject the largest amounts are excluded ($-0.056$, $p=0.386$), although the association remains in a rank-based specification ($p=0.019$) (Online Appendix, Section J). Second, 47 of the 99 judges accept every offer they receive; for these judges the amount accepted coincides with the amount offered, so that the two cannot be separated. Punishment inflicted remains positive and significant.

Panel B expresses the same margin as a share. The bribe rejection share is negatively associated with perceived corruption: the coefficient is $-0.616$ ($p=0.050$) in the baseline specification and $-0.584$ ($p=0.043$) after adding background characteristics, so that a 10-percentage-point increase in the share of positive bribes rejected is associated with approximately 0.062 points lower reported corruption. Alternative measures based on the number of positive bribes rejected and on whether a judge ever rejects a positive offer yield the same negative sign (Online Appendix, Section E). Panels A and B describe the same behavior, the rejection of bribes, measured in two ways, and should be read as one piece of evidence rather than two. This evidence is less robust than the evidence on punishment. The rejection-share coefficient is no longer significant at the 5 percent level with attitudinal controls, with session fixed effects, or when the first period is excluded (Online Appendix, Sections C, I, and L). Moreover, among the 52 judges who reject at least one offer, the share rejected is unrelated to perceived corruption ($-0.194$, $p=0.594$; Online Appendix, Section J): the relationship reflects mainly the difference between judges who never reject an offer and judges who do.

\subsection{Robustness}
\label{subsec:robustness}

The main cross-role results change little when the specification is modified. With the broader set of attitudinal controls, citizens' behavioral coefficients remain close to zero and the joint null is not rejected ($p=0.999$), whereas the judge behavioral measures remain jointly significant ($p=0.023$) (Online Appendix, Section C). Ordered-logit estimates, which account for the ordinal nature of the corruption-perception measure, preserve the signs of all judge coefficients; punishment inflicted remains statistically significant ($p=0.015$), while bribes accepted and the rejection share are significant at the 10 percent level ($p=0.085$ and $p=0.073$) (Online Appendix, Section F).

The cross-role pattern also persists under different treatments of the repeated structure of the data. When each period enters as a separate observation and standard errors are clustered at the participant level, citizens' behavioral measures remain jointly unrelated to perceived corruption ($p=0.997$ and $p=0.974$), whereas the judge behavioral measures remain jointly significant ($p=0.015$ and $p=0.007$) (Online Appendix, Section G). Clustering at the fixed-group level leaves the citizen results unchanged (Online Appendix, Section H). Replacing experimental-condition indicators with session fixed effects preserves the null results among citizens and the joint significance of the judge measures ($p=0.005$ and $p=0.006$); punishment inflicted remains significant, while the coefficient on bribes accepted is estimated less precisely ($p=0.121$ and $p=0.062$). The rejection-share coefficient remains negative and similar in magnitude under session fixed effects but is no longer significant at the 5 percent level ($p=0.087$ and $p=0.101$) (Online Appendix, Section I).

Finally, the results do not depend on the definition of the estimation sample. Excluding the first period, or retaining for judges the periods of the Exit condition in which no interaction takes place, leaves the citizen results and the joint significance of the judge measures unchanged; the joint test of the difference between roles ranges from $p=0.029$ to $p=0.049$ across these definitions. The rejection-share coefficient is smaller and less precisely estimated when the first period is excluded ($-0.500$, $p=0.098$) (Online Appendix, Section L). For citizens, the outcome of their own bribe offers is likewise unrelated to perceived corruption (Online Appendix, Section M).

\section{Discussion and Conclusion}
\label{sec:conclusion}

This paper examines whether a general survey-style assessment of corruption corresponds to corruption-related behavior observed directly under incentives and whether that correspondence varies with the position individuals occupy within the same corrupt interaction. The results differ across positions. Among citizens, general corruption perceptions show no systematic relationship with bribes offered, contributions to the common project, punishment received, or the outcome of their own bribe offers. Among judges, accepted bribery and punishment inflicted are jointly associated with higher perceived corruption, and perceived corruption rises gradually with the amount of bribery accepted, from 2.61 among the third of judges who accept the least to 2.88 among the third who accept the most. Hypothesis 1 therefore receives support in the judge position but not in the citizen position. A direct comparison supports Hypothesis 2: when behavior is expressed in standard-deviation units, its association with perceived corruption is larger for judges than for citizens (joint test, $p=0.029$). The evidence for this difference is moderate rather than strong. It is driven mainly by punishment, and it is weaker when behavior is measured in original units, which are not directly comparable across roles. Among the judge measures, punishment inflicted is the most robust: its association with perceived corruption remains significant when any single judge is excluded. The evidence on bribe rejection points in the same direction, with judges who reject more of the offers they receive reporting lower corruption, but it is less precise.

The citizen results fit within the broader evidence showing that personal experience and general corruption perceptions do not necessarily move closely together, although the strength of this correspondence varies substantially across settings \citep{WeberAbramo2008,RoseMishler2010,Charron2016,GutmannPadovanoVoigt2020}. Our results add a behavioral dimension to this heterogeneous evidence. Even when corruption-related choices are observed directly rather than reconstructed from self-reports, correspondence with a general corruption assessment is not automatic. The pattern differs across positions within the same experimental environment.

Our findings also complement the small literature combining survey information with experimentally observed corruption \citep{CamposOrtiz2011,BuonannoPucaSbriglia2026}. Rather than asking whether prior experience, attitudes, or contextual characteristics predict experimental behavior, we examine whether a general corruption assessment corresponds to the same individual's revealed behavior and whether that correspondence differs across positions within the interaction.

The difference between positions is consistent with evidence that the informational content of corruption perceptions depends on what individuals are able to observe. \citet{Olken2009} shows that villagers' assessments of corruption in public projects become more informative when the underlying activity is more observable. In our setting, this distinction is not reducible to a simple information/no-information contrast: citizens' access to others' behavior varies across experimental conditions, whereas judges systematically observe all citizens' behavior and exercise direct authority over bribe acceptance and punishment. The judge results give some indication that exposure is not the whole story: the amount of bribery offered to a judge is not by itself associated with the judge's perception, while the judge's own decisions, punishment in particular, are. This indication is limited, however. Almost half of the judges accept every offer, so that for them the amount accepted is the amount offered, and the design does not vary exposure independently of decision authority. The pattern is therefore consistent with the joint relevance of observational scope and decision authority, in line with related evidence of role-specific responses within corrupt interactions \citep{LevatiNardi2023,BuonannoPucaSbriglia2026,MonteroPossajennikovVerbel2026}, but it does not separate the two.

These findings also matter for the interpretation of survey-based corruption measures. General corruption questions are often treated as comparable assessments of the same underlying phenomenon, even though respondents may occupy very different positions relative to corrupt exchanges. Where feasible, surveys could therefore distinguish not only whether respondents have encountered corruption, but also whether they did so as potential bribe payers, recipients, decision makers, observers, or affected third parties. Such information may help explain why similar general assessments carry different behavioral content across respondents.

Two features of the design delimit the scope of this interpretation. First, the perception item concerns corruption in the country rather than corruption within the experimental group. The analysis therefore does not test whether participants accurately estimate the amount of corruption generated inside the laboratory; it asks instead whether a general survey-style assessment corresponds to behavior revealed in a controlled corruption interaction. Second, the perception measure is elicited after the experimental tasks. The observed associations therefore do not identify whether pre-existing perceptions shape subsequent behavior, experimental experience contributes to later perceptions, or both reflect common individual characteristics. The contribution of the design is instead comparative: all participants answer the same general question after experiencing the same underlying corruption environment from systematically different positions.

Overall, the results suggest that the behavioral content of corruption perceptions depends partly on the position from which corruption is encountered. Accounting for that position can help explain why similar general assessments map differently onto individual behavior and can improve the interpretation of perception-based measures of corruption.

% ------------------------------------------------
% Declarations
% ------------------------------------------------

\section*{CRediT authorship contribution statement}

\textbf{Francesca Pancotto:} Conceptualization, Methodology, Formal analysis, Investigation, Data curation, Writing -- original draft, Writing -- review \& editing.

\textbf{Enrico D'Ecclesiis:} Conceptualization, Methodology, Formal analysis, Investigation, Data curation, Writing -- original draft, Writing -- review \& editing.

\section*{Declaration of competing interest}

The authors declare that they have no known competing financial interests or personal relationships that could have appeared to influence the work reported in this paper.

\section*{Funding}

The experiment was funded by the VERA project, Università Ca' Foscari Venezia. F.P. was funded by FAR2024 UNIMORE (CUP E83D2400063000). E.D. was funded by FAR2025 DEMB UNIMORE. The funding sources had no involvement in the study design, in the collection, analysis and interpretation of the data, in the writing of the article, or in the decision to submit it for publication.

\section*{Appendix A. Supplementary material}

Supplementary material related to this article is provided in the Online Appendix. \textit{Online Appendix:} definitions of all measures and variables (Section~A); descriptive statistics (Section~B); complete regression tables for the specifications reported in the paper (Section~C); additional specifications and robustness analyses (Sections~D--M); and a summary of the experimental instructions for the bribery game (Section~N).

\section*{Acknowledgments}

We thank the staff of VERALabEx for assistance with the experimental sessions.

\section*{Declaration of generative AI and AI-assisted technologies in the manuscript preparation process}

During the preparation of this work, the authors used ChatGPT (OpenAI) and Claude (Anthropic). The tools were used to improve the language, structure, and presentation of the manuscript; to obtain a critical review of earlier drafts; and to assist with the replication of the statistical analysis, the generation of the tables and of Figure~\ref{fig:perception_tercile}, and the drafting of revisions to the text. All estimates reported in the paper and in the Online Appendix were re-estimated and verified by the authors. After using these tools, the authors reviewed and edited the content as needed and take full responsibility for the content of the published article.

% ------------------------------------------------
% References
% ------------------------------------------------

\bibliographystyle{apalike}
\bibliography{main}

\end{document}

\maketitle

\tableofcontents

\newpage

\appendix

\numberwithin{table}{section}
\numberwithin{figure}{section}

% ================================================================
% A. MEASURES AND VARIABLE CONSTRUCTION
% ================================================================

\section{Measures and Variable Construction}
\label{app:measures}

This section provides additional details on the construction and coding of the outcome, experimental behavioral measures, and individual characteristics used in the empirical analysis.

\subsection{Corruption Perception}
\label{app:corruption_measure}

The outcome variable, \textit{corruption perception}, is obtained from the final questionnaire administered after completion of the experimental tasks. Participants were asked: ``How widespread do you think corruption is in this country?'' Responses were coded on a four-category scale: 1 if hardly any of the public offices are corrupt, 2 if few public officials are corrupt, 3 if most public officials are corrupt, and 4 if almost all public officials are corrupt. Higher values therefore indicate a greater perceived prevalence of corruption. Although the response scale ranges from 1 to 4, the observed responses in the estimation sample lie between 2 and 4. The measure refers to corruption in the country rather than to corruption within the experimental group and is therefore interpreted as a general stated corruption perception rather than as an estimate of the amount of corruption observed during the experiment.

\subsection{Citizens' Behavioral Measures}
\label{app:citizen_measures}

The citizen analysis uses three behavioral measures observed at the period level. \textit{Bribe offered} records the number of points that the citizen transfers to the judge during the bribery stage and ranges from 0 to 5. A value of zero indicates that no transfer is offered. \textit{Contribution} records the number of points allocated by the citizen to the common project and ranges from 0 to 20. \textit{Punishment received} records the effective reduction in the citizen's earnings generated by the judge's punishment decision. Since each deduction point assigned by the judge reduces the targeted citizen's earnings by three points, this measure equals three times the number of deduction points received and ranges from 0 to 30. In the Exit condition, observations corresponding to the special non-participation code used when a citizen does not participate in the interaction are treated as missing rather than as zero punishment. The main analysis enters these behavioral measures jointly, while the corresponding separate specifications are reported as supplementary evidence in Section~\ref{app:separate}. Complete versions of the main joint specifications are reported in Section~\ref{app:complete}.

\subsection{Judges' Behavioral Measures}
\label{app:judge_measures}

The judge analysis uses two main period-level measures corresponding to decisions associated with the judge's role. \textit{Bribes accepted} records the total amount of bribery accepted by the judge in each period and ranges from 0 to 15. Transfers rejected by the judge are returned to the corresponding citizens and therefore do not contribute to this measure. \textit{Punishment inflicted} records the total number of deduction points allocated by the judge to citizens during the period. These variables capture, respectively, participation in completed bribery transactions and the exercise of punishment within the group. In the Exit condition, periods in which the common project is not implemented because two or more citizens withdraw involve no acceptance, rejection, or punishment decision by the judge and are treated as missing rather than as zero bribes accepted and zero punishment. The main analysis enters these behavioral measures jointly, while the corresponding separate specifications are reported as supplementary evidence in Section~\ref{app:separate}. Complete versions of the main joint specifications are reported in Section~\ref{app:complete}. In the additional analysis reported in Section~\ref{app:bribe_opportunities}, we also use \textit{total bribes offered}, defined as the sum of the bribe offers made by the three citizens to the judge in a given period. This measure ranges from 0 to 15 and captures the amount of bribery opportunities received by the judge independently of whether those offers are subsequently accepted or rejected.

\subsection{Measures of Revealed Resistance to Bribery}
\label{app:rejection_measures}

We construct three judge-level measures of revealed resistance to bribery using positive bribe offers only. Zero transfers are excluded because they do not provide the judge with an opportunity to reveal whether an offered bribe would be accepted or rejected. The main measure, \textit{bribe rejection share}, is defined as the number of positive bribe offers rejected by the judge divided by the total number of positive bribe offers received over all periods of the experiment. It therefore ranges from 0 to 1, with higher values indicating a greater revealed propensity to refuse available bribery opportunities. The first alternative measure, \textit{bribe rejection count}, records the total number of positive bribe offers rejected by the judge. The second, \textit{ever rejected}, is an indicator equal to one if the judge rejects at least one positive bribe offer during the experiment and zero otherwise. These measures describe behavior revealed during the experiment and are not interpreted as measures of ex ante or intrinsic honesty.

\subsection{Participant-Level Aggregation and Estimation Sample}
\label{app:aggregation}

Corruption perception is observed once per participant, whereas the behavioral measures are observed in every period. The main analysis therefore uses one observation per participant, obtained by averaging each period-level behavioral measure within participant over all periods of the bribery game in which the participant takes part in the interaction. The 297 citizens generate 3,885 citizen-period observations. Of these, 242 are periods in which the citizen chooses the outside option in the Exit condition and 47 are periods in which the citizen is the only one not withdrawing, so that the common project is not implemented; in both cases no contribution, transfer, or punishment takes place and the period is treated as missing. This leaves 3,596 citizen-period observations for 295 citizens, since two citizens choose the outside option in every period. The 99 judges generate 1,295 judge-period observations, of which 55 (involving 16 judges in the Exit condition) are periods in which two or more citizens withdraw and no interaction takes place; excluding them leaves 1,240 judge-period observations for 99 judges. Specifications based on the period-level observations are reported in Section~\ref{app:period}, and specifications based on alternative definitions of the sample, including the exclusion of the first period, in Section~\ref{app:sample_definition}.

\subsection{Background and Attitudinal Measures}
\label{app:control_measures}

The empirical analysis uses two sets of individual characteristics collected in the final questionnaire. The first set consists of background characteristics: age, gender, association membership, voting status, and blood-donation status. \textit{Age} is entered as a continuous variable using reported age. \textit{Gender} is self-reported and categorical; the observed response categories are Female, Male, Not answering, and, where observed, Not identified. Female is the reference category in the regressions. \textit{Association membership} distinguishes participants reporting membership in an association from those reporting no membership, with No as the reference category. \textit{Voting status} distinguishes Yes, No, and I can't, with Yes as the reference category. \textit{Blood-donation status} distinguishes Yes, No, and I didn't meet the requirements / age, with Yes as the reference category.

The second set consists of attitudinal measures used in robustness specifications. \textit{Trust} measures generalized trust, while \textit{Benefits}, \textit{Fare}, \textit{Tax}, and \textit{Bribe} record respondents' stated assessments of the justifiability of claiming undeserved benefits, fare evasion, tax evasion, and accepting a bribe, respectively. These variables are recorded on their original six-point scales and enter the regressions linearly. For generalized trust, the scale ranges from 1 (``You should be very careful'') to 6 (``You can trust people''), so that higher values indicate greater generalized trust. For each of the four justifiability measures, the scale ranges from 1 (``Never justified'') to 6 (``Always justified''), so that higher values indicate greater stated justifiability of the corresponding behavior. These attitudinal measures are added to assess whether the main behavioral relationships persist after accounting for a broader set of individual attitudes.

All regressions additionally include indicators for the experimental condition unless otherwise specified. FULL (Full Information) is the omitted reference category; the estimated indicators correspond to PINFO (Control), PENDO (Transparency), and PENDO\_EXIT (Exit).

% ================================================================
% B. DESCRIPTIVE STATISTICS
% ================================================================

\section{Descriptive Statistics}
\label{app:descriptives}

This section reports descriptive statistics for the variables used in the empirical analysis. Table~\ref{tab:descriptive_sample} summarizes the corruption-perception outcome and the continuous or ordinal individual characteristics for the full sample of 396 participants. Table~\ref{tab:descriptive_behavior} reports the experimental behavioral measures, indicating the institutional role and level of observation to which each measure applies. Table~\ref{tab:descriptive_categorical} reports the complete distributions of the categorical characteristics and experimental conditions in the full sample. Table~\ref{tab:perception_by_tercile} reports average perceived corruption by level of bribery, corresponding to Figure~1 of the paper.

\subsection{Overall Sample Characteristics}
\label{app:sample_descriptives}

\begin{table}[H]
\centering
\caption{Descriptive statistics: Overall sample}
\label{tab:descriptive_sample}

\begin{threeparttable}
\small
\begin{tabular}{lrrrrr}
\toprule
Variable & $N$ & Mean & SD & Min & Max \\
\midrule

Corruption perception
& 396 & 2.788 & 0.561 & 2 & 4 \\

Age
& 396 & 21.707 & 3.723 & 18 & 60 \\

Generalized trust
& 396 & 3.018 & 1.148 & 1 & 6 \\

Justifiability: Undeserved benefits
& 396 & 1.813 & 1.048 & 1 & 6 \\

Justifiability: Fare evasion
& 396 & 2.556 & 1.333 & 1 & 6 \\

Justifiability: Tax evasion
& 396 & 1.826 & 1.344 & 1 & 6 \\

Justifiability: Accepting a bribe
& 396 & 2.205 & 1.425 & 1 & 6 \\

\bottomrule
\end{tabular}
\end{threeparttable}

\vspace{0.15cm}

\parbox{\textwidth}{\footnotesize\textit{Notes:} Statistics are calculated using one observation per participant. Corruption perception ranges from 1 to 4, with higher values indicating greater perceived corruption; observed responses in the sample range from 2 to 4. Generalized trust and the four justifiability measures are reported on their original six-point scales.}

\end{table}

The sample consists of 396 participants, of whom 297 were assigned the citizen role and 99 the judge role. Average reported corruption perception is 2.788 on the four-category scale. Participants are relatively young on average, with a mean age of 21.71 years, and the observed age range extends from 18 to 60. The attitudinal measures display variation across their respective six-point scales, with average generalized trust of 3.02 and relatively low average stated justifiability of the rule-breaking behaviors included as controls.

\subsection{Experimental Behavioral Measures}
\label{app:behavior_descriptives}

\begin{table}[H]
\centering
\caption{Descriptive statistics: Experimental behavioral measures}
\label{tab:descriptive_behavior}

\begin{threeparttable}
\small
\begin{tabular}{llrrrrr}
\toprule
Variable & Role / level & $N$ & Mean & SD & Min & Max \\
\midrule

Bribe offered
& Citizen-period
& 3,596 & 2.085 & 1.633 & 0 & 5 \\

Contribution
& Citizen-period
& 3,596 & 9.970 & 5.949 & 0 & 20 \\

Punishment received
& Citizen-period
& 3,596 & 2.299 & 4.505 & 0 & 30 \\

\addlinespace

Bribes accepted
& Judge-period
& 1,240 & 5.408 & 3.540 & 0 & 15 \\

Total bribes offered
& Judge-period
& 1,240 & 6.047 & 3.468 & 0 & 15 \\

Punishment inflicted
& Judge-period
& 1,240 & 2.223 & 3.177 & 0 & 15 \\

\addlinespace

Bribe offered, participant mean
& Citizen
& 295 & 2.053 & 1.258 & 0 & 5.000 \\

Contribution, participant mean
& Citizen
& 295 & 9.827 & 4.701 & 0 & 20.000 \\

Punishment received, participant mean
& Citizen
& 295 & 2.361 & 2.901 & 0 & 20.250 \\

Bribes accepted, participant mean
& Judge
& 99 & 5.280 & 2.802 & 0.500 & 12.643 \\

Total bribes offered, participant mean
& Judge
& 99 & 5.901 & 2.844 & 0.909 & 13.000 \\

Punishment inflicted, participant mean
& Judge
& 99 & 2.235 & 2.306 & 0 & 15.000 \\

\addlinespace

Bribe rejection share
& Judge
& 99 & 0.109 & 0.178 & 0 & 0.854 \\

Bribe rejection count
& Judge
& 99 & 2.818 & 5.122 & 0 & 35 \\

Ever rejected a positive bribe
& Judge
& 99 & 0.525 & 0.502 & 0 & 1 \\

\bottomrule
\end{tabular}
\end{threeparttable}

\vspace{0.15cm}

\parbox{\textwidth}{\footnotesize\textit{Notes:} Period-level statistics are calculated over all periods of the bribery game in which the participant takes part in the interaction. Citizen-periods in which the citizen chooses the outside option, and periods of the Exit condition in which the common project is not implemented because two or more citizens withdraw, are treated as missing for both roles. Participant means are the within-participant averages used in the participant-level specifications. Punishment received is the effective reduction in the citizen's earnings (three points per deduction point). Total bribes offered is the sum of the bribe offers made by the citizens to the judge in the period. Bribe rejection share, bribe rejection count, and ever rejected are judge-level measures constructed from positive bribe offers: the proportion of positive offers rejected, the total number of positive offers rejected, and an indicator equal to one if the judge rejects at least one positive offer.}

\end{table}

Citizens offer an average of 2.09 of the five points available for transfer to the judge and contribute an average of 9.97 of the 20 points available for the common project. Average effective punishment received is 2.30 points. Judges receive an average of 6.05 points in total bribe offers per period, accept an average of 5.41 points, and allocate an average of 2.22 deduction points. At the judge level, the average rejection share is 0.109, corresponding to approximately 11 percent of positive offers rejected. Judges reject an average of 2.82 positive offers over the experiment, and 52.5 percent reject at least one positive bribe offer. The rejection measures are strongly right-skewed: almost half of the judges never reject a positive offer, while a small number reject a large share of the offers they receive.

\subsection{Categorical Characteristics and Experimental Conditions}
\label{app:categorical_descriptives}

\begin{table}[H]
\centering
\caption{Distribution of categorical characteristics and experimental conditions}
\label{tab:descriptive_categorical}

\begin{threeparttable}
\small
\begin{tabular}{llrr}
\toprule
Variable & Category & $N$ & Percent \\
\midrule

\textit{Experimental role}
& Citizen & 297 & 75.00 \\
& Judge & 99 & 25.00 \\

\addlinespace
\textit{Experimental condition}
& FULL & 36 & 9.09 \\
& PENDO & 124 & 31.31 \\
& PENDO\_EXIT & 116 & 29.29 \\
& PINFO & 120 & 30.30 \\

\addlinespace
\textit{Gender}
& Female & 236 & 59.60 \\
& Male & 150 & 37.88 \\
& Not answering & 8 & 2.02 \\
& Not identified & 2 & 0.51 \\

\addlinespace
\textit{Association membership}
& No & 105 & 26.52 \\
& Yes & 291 & 73.48 \\

\addlinespace
\textit{Voting status}
& Yes & 308 & 77.78 \\
& No & 45 & 11.36 \\
& I can't & 43 & 10.86 \\

\addlinespace
\textit{Blood-donation status}
& Yes & 57 & 14.39 \\
& No & 262 & 66.16 \\
& Did not meet requirements / age & 77 & 19.44 \\

\bottomrule
\end{tabular}
\end{threeparttable}

\vspace{0.15cm}

\parbox{\textwidth}{\footnotesize\textit{Notes:} The table reports frequencies and percentages using one observation per participant ($N=396$). FULL denotes Full Information, PINFO denotes Control, PENDO denotes Transparency, and PENDO\_EXIT denotes Exit. All observed response categories are retained in the empirical analysis. In the regression specifications, Female is the reference category for gender, No for association membership, Yes for voting status, Yes for blood-donation status, and FULL for the experimental condition.}

\end{table}

The experimental role distribution follows directly from the four-person group structure, with three citizens and one judge in each of the 99 groups. Thirty-six participants were assigned to Full Information, 124 to Transparency, 116 to Exit, and 120 to Control. Female participants account for 59.6 percent of the sample, 73.5 percent report association membership, and 77.8 percent report having voted. All observed response categories, including the less frequent non-response categories for gender, are retained in the regression analysis.

\subsection{Perceived Corruption by Level of Bribery}
\label{app:perception_tercile}

Table~\ref{tab:perception_by_tercile} reports the group means displayed in Figure~1 of the paper. Participants in each role are divided into three groups of equal size according to their average bribery over all periods of the bribery game in which they take part in the interaction: the average bribe offered for citizens and the average amount of bribes accepted per period for judges.

\begin{table}[H]
\centering
\caption{Perceived corruption by third of own bribery}
\label{tab:perception_by_tercile}

\begin{threeparttable}
\small
\begin{tabular}{lcrcccc}
\toprule
 & Range of & & \multicolumn{2}{c}{Perceived corruption} & \multicolumn{2}{c}{Share answering (\%)} \\
\cmidrule(lr){4-5}\cmidrule(lr){6-7}
Group & grouping variable & $N$ & Mean & 95\% CI & Few & Almost all \\
\midrule
\multicolumn{7}{l}{\textit{Panel A: Citizens, by average bribe offered}} \\
\quad Low third & 0.00--1.38 & 99 & 2.778 & [2.662,\ 2.894] & 30.3 & 8.1 \\
\quad Middle third & 1.42--2.58 & 98 & 2.857 & [2.749,\ 2.965] & 22.4 & 8.2 \\
\quad High third & 2.58--5.00 & 98 & 2.765 & [2.658,\ 2.872] & 28.6 & 5.1 \\
\addlinespace
\multicolumn{7}{l}{\textit{Panel B: Judges, by average bribes accepted}} \\
\quad Low third & 0.50--4.00 & 33 & 2.606 & [2.390,\ 2.822] & 45.5 & 6.1 \\
\quad Middle third & 4.08--6.46 & 33 & 2.727 & [2.544,\ 2.911] & 30.3 & 3.0 \\
\quad High third & 6.50--12.64 & 33 & 2.879 & [2.666,\ 3.091] & 24.2 & 12.1 \\
\bottomrule
\end{tabular}
\end{threeparttable}

\vspace{0.15cm}

\parbox{\textwidth}{\footnotesize\textit{Notes:} One observation per participant. Perceived corruption is coded from 1 to 4; the observed responses lie between 2 and 4. The range of the grouping variable is in points per period (0 to 5 for citizens, 0 to 15 for judges); groups are formed on ranks, so that participants with the same value at a boundary can fall in adjacent groups. ``Few'' and ``Almost all'' are the shares of participants answering that few public officials are corrupt (category 2) and that almost all public officials are corrupt (category 4). Confidence intervals are based on the $t$ distribution within each group and do not adjust for the experimental condition.}

\end{table}

Among citizens, perceived corruption does not vary across the three groups: the Spearman rank correlation between group and perception is $-0.004$ ($p=0.939$), and the difference between the highest and the lowest third is $-0.012$ ($p=0.875$, two-sample $t$ test with unequal variances). Among judges, the rank correlation is 0.194 ($p=0.055$) and the difference between the highest and the lowest third is 0.273 ($p=0.072$). In a regression of perceived corruption on indicators for the middle and the highest third and experimental-condition indicators, with robust standard errors, the coefficient on the middle third is 0.158 (standard error 0.140, $p=0.264$) and the coefficient on the highest third is 0.302 (standard error 0.156, $p=0.055$); the two indicators are not jointly significant ($p=0.156$). When a quadratic term in the average amount of bribes accepted is added to the linear specification with experimental-condition indicators, it is not statistically significant ($p=0.239$). The grouping into thirds is descriptive; the corresponding tests are those of the participant-level specifications reported in Section~\ref{app:complete}.

% ================================================================
% C. COMPLETE MAIN SPECIFICATIONS
% ================================================================

\section[Complete Main Specifications and Attitudinal-Control Robustness]{Complete Main Specifications and\\ Attitudinal-Control Robustness}
\label{app:complete}

This section reports complete versions of the main behavioral specifications and the corresponding robustness specifications including the broader set of attitudinal measures. All specifications use one observation per participant. For citizens and judges, columns (1) and (2) reproduce the joint baseline and background-controlled models underlying the main cross-role results while reporting all estimated covariates. Column (3) additionally includes generalized trust and the stated justifiability of claiming undeserved benefits, fare evasion, tax evasion, and accepting a bribe.

\subsection{Citizens}
\label{app:complete_citizens}

Table~\ref{tab:complete_citizens} reports the complete versions of the citizen specifications presented in the main text. The coefficients on bribes offered, contributions, and punishment received remain small and statistically indistinguishable from zero throughout the three specifications, and the joint null is not rejected ($p=0.907$, $p=0.962$, and $p=0.999$). The complete estimates therefore confirm that the absence of behavioral correspondence among citizens is unaffected by the inclusion of the observed background and attitudinal characteristics.

\begin{longtable}{p{6.8cm}ccc}
\caption{Complete specifications: Citizens}
\label{tab:complete_citizens}\\
\toprule
& (1) & (2) & (3) \\
& Baseline & Background & + Attitudinal controls \\
\midrule
\endfirsthead

\multicolumn{4}{l}{\small\textit{Table \thetable\ continued from previous page}}\\
\toprule
& (1) & (2) & (3) \\
& Baseline & Background & + Attitudinal controls \\
\midrule
\endhead

\midrule
\multicolumn{4}{r}{\small\textit{Continued on next page}}\\
\endfoot

\bottomrule
\endlastfoot

Bribe offered
& -0.0101 & 0.0005 & 0.0032 \\
& (0.0292) & (0.0276) & (0.0270) \\

Contribution
& -0.0027 & -0.0034 & -0.0006 \\
& (0.0071) & (0.0067) & (0.0067) \\

Punishment received
& 0.0022 & -0.0002 & 0.0010 \\
& (0.0114) & (0.0102) & (0.0103) \\

\addlinespace
Age
&  & -0.0251$^{**}$ & -0.0209$^{*}$ \\
&  & (0.0109) & (0.0113) \\

Gender: Male
&  & -0.3204$^{***}$ & -0.3454$^{***}$ \\
&  & (0.0686) & (0.0688) \\

Gender: Not answering
&  & -0.1598 & -0.1241 \\
&  & (0.1777) & (0.1774) \\

Association membership: Yes
&  & -0.1820$^{**}$ & -0.1586$^{**}$ \\
&  & (0.0758) & (0.0761) \\

Voting status: No
&  & 0.1163 & 0.0941 \\
&  & (0.0974) & (0.0983) \\

Voting status: I can't
&  & -0.0867 & -0.1003 \\
&  & (0.0989) & (0.1020) \\

Blood-donation status: No
&  & -0.0115 & -0.0239 \\
&  & (0.0869) & (0.0931) \\

Blood-donation status: Did not meet requirements / age
&  & 0.0772 & 0.0448 \\
&  & (0.1000) & (0.1037) \\

\addlinespace
Generalized trust
&  &  & -0.0394 \\
&  &  & (0.0272) \\

Justifiability: Undeserved benefits
&  &  & -0.0379 \\
&  &  & (0.0339) \\

Justifiability: Fare evasion
&  &  & -0.0187 \\
&  &  & (0.0259) \\

Justifiability: Tax evasion
&  &  & 0.0713$^{***}$ \\
&  &  & (0.0265) \\

Justifiability: Accepting a bribe
&  &  & 0.0202 \\
&  &  & (0.0238) \\

\addlinespace
PENDO
& -0.1172 & -0.1386 & -0.1619$^{*}$ \\
& (0.1085) & (0.0965) & (0.0926) \\

PENDO\_EXIT
& -0.2213$^{**}$ & -0.2847$^{***}$ & -0.2885$^{***}$ \\
& (0.1103) & (0.1002) & (0.0952) \\

PINFO
& -0.0773 & -0.1041 & -0.1085 \\
& (0.1094) & (0.0995) & (0.0972) \\

\addlinespace
Constant
& 2.9666$^{***}$ & 3.7729$^{***}$ & 3.7242$^{***}$ \\
& (0.1208) & (0.2828) & (0.2979) \\

\midrule
Participants
& 295 & 295 & 295 \\

$R^{2}$
& 0.0169 & 0.1233 & 0.1628 \\

\end{longtable}

{\footnotesize\noindent\textit{Notes:} OLS estimates using one observation per citizen. The dependent variable is perceived corruption, coded from 1 to 4, with higher values indicating greater perceived corruption. Behavioral measures are averaged within participant over all periods of the bribery game in which the participant takes part in the interaction. Periods of the Exit condition in which the common project is not implemented because two or more citizens withdraw are treated as missing. Robust standard errors are reported in parentheses. Female is the reference category for gender, No for association membership, Yes for voting status, Yes for blood-donation status, and FULL (Full Information) for the experimental condition. PINFO denotes Control, PENDO denotes Transparency, and PENDO\_EXIT denotes Exit. Generalized trust and the four justifiability measures enter on their original six-point scales. $^{*}p<0.10$, $^{**}p<0.05$, $^{***}p<0.01$.\par}

\hspace{1cm}

\subsection{Judges}
\label{app:complete_judges}

Table~\ref{tab:complete_judges} reports the complete versions of the judge specifications presented in the main text. The coefficients on bribes accepted and punishment inflicted are positive and stable in magnitude across specifications. Punishment inflicted is statistically significant throughout ($p=0.013$, $p=0.016$, and $p=0.022$). Bribes accepted is estimated less precisely ($p=0.089$, $p=0.036$, and $p=0.055$). The two measures are jointly significant in all three specifications ($p=0.016$, $p=0.014$, and $p=0.023$). The baseline results do not depend on any single judge: when each judge is excluded in turn, the $p$-value of punishment inflicted ranges from 0.004 to 0.042 and the $p$-value of the joint test from 0.002 to 0.046. The coefficient on the category Gender: Not answering is identified by a single judge and should not be interpreted.

\begin{longtable}{p{6.8cm}ccc}
\caption{Complete specifications: Judges}
\label{tab:complete_judges}\\
\toprule
& (1) & (2) & (3) \\
& Baseline & Background & + Attitudinal controls \\
\midrule
\endfirsthead

\multicolumn{4}{l}{\small\textit{Table \thetable\ continued from previous page}}\\
\toprule
& (1) & (2) & (3) \\
& Baseline & Background & + Attitudinal controls \\
\midrule
\endhead

\midrule
\multicolumn{4}{r}{\small\textit{Continued on next page}}\\
\endfoot

\bottomrule
\endlastfoot

Bribes accepted
& 0.0412$^{*}$ & 0.0503$^{**}$ & 0.0474$^{*}$ \\
& (0.0239) & (0.0235) & (0.0243) \\

Punishment inflicted
& 0.0687$^{**}$ & 0.0728$^{**}$ & 0.0722$^{**}$ \\
& (0.0273) & (0.0297) & (0.0309) \\

\addlinespace
Age
&  & -0.0219$^{**}$ & -0.0179$^{*}$ \\
&  & (0.0091) & (0.0102) \\

Gender: Male
&  & -0.2962$^{**}$ & -0.3203$^{**}$ \\
&  & (0.1299) & (0.1244) \\

Gender: Not answering
&  & 0.3297$^{**}$ & 0.3225 \\
&  & (0.1571) & (0.2191) \\

Gender: Not identified
&  & -0.0067 & 0.0066 \\
&  & (0.1456) & (0.1721) \\

Association membership: Yes
&  & -0.2467$^{*}$ & -0.2506$^{*}$ \\
&  & (0.1365) & (0.1294) \\

Voting status: No
&  & -0.1245 & -0.1483 \\
&  & (0.1626) & (0.1611) \\

Voting status: I can't
&  & 0.1752 & 0.1990 \\
&  & (0.2362) & (0.2398) \\

Blood-donation status: No
&  & 0.1127 & 0.0832 \\
&  & (0.1362) & (0.1339) \\

Blood-donation status: Did not meet requirements / age
&  & 0.1324 & 0.1139 \\
&  & (0.1698) & (0.1772) \\

\addlinespace
Generalized trust
&  &  & -0.0441 \\
&  &  & (0.0657) \\

Justifiability: Undeserved benefits
&  &  & -0.0064 \\
&  &  & (0.0664) \\

Justifiability: Fare evasion
&  &  & 0.0212 \\
&  &  & (0.0534) \\

Justifiability: Tax evasion
&  &  & 0.0140 \\
&  &  & (0.0485) \\

Justifiability: Accepting a bribe
&  &  & 0.0305 \\
&  &  & (0.0596) \\

\addlinespace
PENDO
& -0.0231 & -0.1702 & -0.1608 \\
& (0.2553) & (0.2585) & (0.2679) \\

PENDO\_EXIT
& 0.0121 & -0.1432 & -0.1610 \\
& (0.2499) & (0.2681) & (0.3182) \\

PINFO
& -0.1765 & -0.2888 & -0.2794 \\
& (0.2539) & (0.2661) & (0.2783) \\

\addlinespace
Constant
& 2.4238$^{***}$ & 3.1892$^{***}$ & 3.1458$^{***}$ \\
& (0.2557) & (0.4707) & (0.5854) \\

\midrule
Participants
& 99 & 99 & 99 \\

$R^{2}$
& 0.1176 & 0.2351 & 0.2516 \\

\end{longtable}

{\footnotesize\noindent\textit{Notes:} OLS estimates using one observation per judge. The dependent variable is perceived corruption, coded from 1 to 4, with higher values indicating greater perceived corruption. Behavioral measures are averaged within participant over all periods of the bribery game in which the participant takes part in the interaction. Periods of the Exit condition in which the common project is not implemented because two or more citizens withdraw are treated as missing. Robust standard errors are reported in parentheses. Female is the reference category for gender, No for association membership, Yes for voting status, Yes for blood-donation status, and FULL (Full Information) for the experimental condition. PINFO denotes Control, PENDO denotes Transparency, and PENDO\_EXIT denotes Exit. Generalized trust and the four justifiability measures enter on their original six-point scales. $^{*}p<0.10$, $^{**}p<0.05$, $^{***}p<0.01$.\par}

\hspace{1cm}

\subsection{Judges' Revealed Resistance to Bribery}
\label{app:complete_rejection}

Table~\ref{tab:complete_rejection} reports the complete specifications for the main measure of judges' revealed resistance to bribery. The association between the share of positive bribe offers rejected and perceived corruption is negative throughout. The coefficient is $-0.616$ in the baseline specification ($p=0.050$), $-0.584$ after background characteristics are included ($p=0.043$), and $-0.501$ after the additional attitudinal measures are introduced ($p=0.104$).

\begin{longtable}{p{6.8cm}ccc}
\caption{Complete specifications: Judges' revealed resistance to bribery}
\label{tab:complete_rejection}\\
\toprule
& (1) & (2) & (3) \\
& Baseline & Background & + Attitudinal controls \\
\midrule
\endfirsthead

\multicolumn{4}{l}{\small\textit{Table \thetable\ continued from previous page}}\\
\toprule
& (1) & (2) & (3) \\
& Baseline & Background & + Attitudinal controls \\
\midrule
\endhead

\midrule
\multicolumn{4}{r}{\small\textit{Continued on next page}}\\
\endfoot

\bottomrule
\endlastfoot

Bribe rejection share
& -0.6156$^{*}$ & -0.5836$^{**}$ & -0.5006 \\
& (0.3100) & (0.2839) & (0.3046) \\

\addlinespace
Age
&  & -0.0183$^{*}$ & -0.0150 \\
&  & (0.0104) & (0.0114) \\

Gender: Male
&  & -0.2977$^{**}$ & -0.3149$^{**}$ \\
&  & (0.1304) & (0.1263) \\

Gender: Not answering
&  & 0.1760 & 0.2019 \\
&  & (0.1805) & (0.2319) \\

Gender: Not identified
&  & 0.1871 & 0.2187 \\
&  & (0.1879) & (0.2038) \\

Association membership: Yes
&  & -0.1355 & -0.1442 \\
&  & (0.1438) & (0.1407) \\

Voting status: No
&  & -0.0527 & -0.0815 \\
&  & (0.1893) & (0.1925) \\

Voting status: I can't
&  & 0.1802 & 0.2053 \\
&  & (0.2327) & (0.2317) \\

Blood-donation status: No
&  & 0.1139 & 0.0843 \\
&  & (0.1611) & (0.1603) \\

Blood-donation status: Did not meet requirements / age
&  & 0.0578 & 0.0499 \\
&  & (0.2137) & (0.2218) \\

\addlinespace
Generalized trust
&  &  & -0.0530 \\
&  &  & (0.0668) \\

Justifiability: Undeserved benefits
&  &  & 0.0172 \\
&  &  & (0.0819) \\

Justifiability: Fare evasion
&  &  & 0.0018 \\
&  &  & (0.0542) \\

Justifiability: Tax evasion
&  &  & 0.0179 \\
&  &  & (0.0540) \\

Justifiability: Accepting a bribe
&  &  & 0.0285 \\
&  &  & (0.0557) \\

\addlinespace
PENDO
& -0.0010 & -0.1433 & -0.1453 \\
& (0.2625) & (0.2889) & (0.2829) \\

PENDO\_EXIT
& -0.0889 & -0.2425 & -0.2773 \\
& (0.2635) & (0.3000) & (0.3226) \\

PINFO
& -0.1679 & -0.2640 & -0.2773 \\
& (0.2705) & (0.2976) & (0.2908) \\

\addlinespace
Constant
& 2.8819$^{***}$ & 3.5280$^{***}$ & 3.5242$^{***}$ \\
& (0.2562) & (0.5344) & (0.5864) \\

\midrule
Participants
& 99 & 99 & 99 \\

$R^{2}$
& 0.0441 & 0.1415 & 0.1584 \\

\end{longtable}

{\footnotesize\noindent\textit{Notes:} OLS estimates using one observation per judge. The dependent variable is perceived corruption, coded from 1 to 4, with higher values indicating greater perceived corruption. The bribe rejection share is the proportion of positive bribe offers rejected by each judge over all periods of the bribery game. Robust standard errors are reported in parentheses. Female is the reference category for gender, No for association membership, Yes for voting status, Yes for blood-donation status, and FULL (Full Information) for the experimental condition. PINFO denotes Control, PENDO denotes Transparency, and PENDO\_EXIT denotes Exit. Generalized trust and the four justifiability measures enter on their original six-point scales. $^{*}p<0.10$, $^{**}p<0.05$, $^{***}p<0.01$.\par}

\hspace{1cm}

% ================================================================
% D. SEPARATE BEHAVIORAL SPECIFICATIONS
% ================================================================

\section{Separate Behavioral Specifications}
\label{app:separate}

The main analysis uses the role-specific behavioral measures jointly as the baseline specification. This section reports the corresponding separate regressions, in which each participant-level behavioral measure enters individually together with the experimental-condition indicators.

\subsection{Citizens}
\label{app:separate_citizens}

Table~\ref{tab:separate_citizens} reports the citizen specifications entering bribes offered, contributions, and punishment received separately. None of the three behavioral measures is systematically associated with perceived corruption. The estimated coefficients are close to zero and statistically insignificant in each specification, consistent with the results from the joint model.

\begin{table}[H]
\centering
\caption{Separate behavioral specifications: Citizens}
\label{tab:separate_citizens}

\begin{threeparttable}
\small
\begin{tabular}{lccc}
\toprule
& (1) & (2) & (3) \\
& Bribe offered & Contribution & Punishment received \\
\midrule

Bribe offered
& -0.0135 &  &  \\
& (0.0271) &  &  \\

Contribution
&  & -0.0038 &  \\
&  & (0.0063) &  \\

Punishment received
&  &  & 0.0033 \\
&  &  & (0.0111) \\

\addlinespace
PENDO
& -0.1195 & -0.1153 & -0.1206 \\
& (0.1075) & (0.1076) & (0.1076) \\

PENDO\_EXIT
& -0.2250$^{**}$ & -0.2164$^{**}$ & -0.2202$^{**}$ \\
& (0.1096) & (0.1086) & (0.1083) \\

PINFO
& -0.0793 & -0.0773 & -0.0821 \\
& (0.1081) & (0.1091) & (0.1079) \\

\addlinespace
Constant
& 2.9545$^{***}$ & 2.9595$^{***}$ & 2.9187$^{***}$ \\
& (0.1098) & (0.1088) & (0.0909) \\

\midrule
Participants
& 295 & 295 & 295 \\

$R^{2}$
& 0.0162 & 0.0163 & 0.0156 \\

\bottomrule
\end{tabular}
\end{threeparttable}

\vspace{0.15cm}

\parbox{\textwidth}{\footnotesize\textit{Notes:} OLS estimates using one observation per citizen. The dependent variable is perceived corruption, coded from 1 to 4, with higher values indicating greater perceived corruption. Each column includes one citizen behavioral measure and indicators for the experimental condition. Behavioral measures are averaged within participant over all periods of the bribery game in which the participant takes part in the interaction. Periods of the Exit condition in which the common project is not implemented because two or more citizens withdraw are treated as missing. Robust standard errors are reported in parentheses. FULL (Full Information) is the omitted experimental condition. PINFO denotes Control, PENDO denotes Transparency, and PENDO\_EXIT denotes Exit. $^{*}p<0.10$, $^{**}p<0.05$, $^{***}p<0.01$.}

\end{table}

\subsection{Judges}
\label{app:separate_judges}

Table~\ref{tab:separate_judges} reports the corresponding specifications for judges. When considered separately, bribes accepted are positively associated with perceived corruption ($p=0.083$), as is punishment inflicted ($p=0.022$). The pattern of the joint specification reported in the main text is therefore already visible when the two dimensions of judge behavior are considered separately.

\begin{table}[H]
\centering
\caption{Separate behavioral specifications: Judges}
\label{tab:separate_judges}

\begin{threeparttable}
\small
\begin{tabular}{lcc}
\toprule
& (1) & (2) \\
& Bribes accepted & Punishment inflicted \\
\midrule

Bribes accepted
& 0.0412$^{*}$ &  \\
& (0.0235) &  \\

Punishment inflicted
&  & 0.0687$^{**}$ \\
&  & (0.0295) \\

\addlinespace
PENDO
& -0.0004 & 0.0060 \\
& (0.2577) & (0.2814) \\

PENDO\_EXIT
& -0.0135 & -0.0281 \\
& (0.2554) & (0.2786) \\

PINFO
& -0.1641 & -0.1235 \\
& (0.2627) & (0.2807) \\

\addlinespace
Constant
& 2.5738$^{***}$ & 2.6277$^{***}$ \\
& (0.2515) & (0.2665) \\

\midrule
Participants
& 99 & 99 \\

$R^{2}$
& 0.0447 & 0.0826 \\

\bottomrule
\end{tabular}
\end{threeparttable}

\vspace{0.15cm}

\parbox{\textwidth}{\footnotesize\textit{Notes:} OLS estimates using one observation per judge. The dependent variable is perceived corruption, coded from 1 to 4, with higher values indicating greater perceived corruption. Each column includes one judge behavioral measure and indicators for the experimental condition. Behavioral measures are averaged within participant over all periods of the bribery game in which the participant takes part in the interaction. Periods of the Exit condition in which the common project is not implemented because two or more citizens withdraw are treated as missing. Robust standard errors are reported in parentheses. FULL (Full Information) is the omitted experimental condition. PINFO denotes Control, PENDO denotes Transparency, and PENDO\_EXIT denotes Exit. $^{*}p<0.10$, $^{**}p<0.05$, $^{***}p<0.01$.}

\end{table}

% ================================================================
% E. ALTERNATIVE MEASURES OF BRIBE REJECTION
% ================================================================

\section{Alternative Measures of Judges' Bribe Rejection}
\label{app:alternative_rejection}

The main analysis measures revealed resistance to bribery using the share of positive bribe offers rejected by each judge. Table~\ref{tab:alternative_rejection} examines whether the same relationship is visible using two alternative judge-level measures: the total number of positive bribe offers rejected and an indicator for whether the judge rejects at least one positive bribe offer during the experiment. As with the main rejection-share specification, each judge contributes one observation.

\begin{table}[H]
\centering
\caption{Alternative measures of judges' bribe rejection}
\label{tab:alternative_rejection}

\begin{threeparttable}
\small
\begin{tabular}{lcc}
\toprule
& (1) & (2) \\
& Rejection count & Ever rejected \\
\midrule

Bribe rejection count
& -0.0231$^{**}$ &  \\
& (0.0091) &  \\

Ever rejected a positive bribe
&  & -0.2200$^{*}$ \\
&  & (0.1162) \\

\addlinespace
PENDO
& -0.0241 & 0.0515 \\
& (0.2669) & (0.2646) \\

PENDO\_EXIT
& -0.1372 & -0.0300 \\
& (0.2726) & (0.2654) \\

PINFO
& -0.1761 & -0.0989 \\
& (0.2730) & (0.2729) \\

\addlinespace
Constant
& 2.9036$^{***}$ & 2.8756$^{***}$ \\
& (0.2627) & (0.2592) \\

\midrule
Observations
& 99 & 99 \\

$R^{2}$
& 0.0494 & 0.0454 \\

\bottomrule
\end{tabular}
\end{threeparttable}

\vspace{0.15cm}

\parbox{\textwidth}{\footnotesize\textit{Notes:} OLS estimates using one observation per judge. The dependent variable is perceived corruption, coded from 1 to 4, with higher values indicating greater perceived corruption. Bribe rejection count is the total number of positive bribe offers rejected by the judge over all periods of the bribery game. Ever rejected equals one if the judge rejects at least one positive bribe offer and zero otherwise. All specifications include indicators for the experimental condition. Robust standard errors are reported in parentheses. FULL (Full Information) is the omitted experimental condition. PINFO denotes Control, PENDO denotes Transparency, and PENDO\_EXIT denotes Exit. $^{*}p<0.10$, $^{**}p<0.05$, $^{***}p<0.01$.}

\end{table}

Both alternatives reproduce the negative pattern found using the rejection share. The total number of positive bribe offers rejected is negatively associated with perceived corruption ($p=0.013$), while judges who reject at least one positive offer also report lower perceived corruption, although this relationship is estimated less precisely ($p=0.061$). The consistency in sign across the three measures indicates that the main result is not specific to the normalization implicit in the rejection-share measure.

% ================================================================
% F. ORDERED-LOGIT SPECIFICATIONS
% ================================================================

\section{Ordered-Logit Specifications}
\label{app:ologit}

Because perceived corruption is measured on an ordered categorical scale, Table~\ref{tab:ologit} re-estimates the main relationships using ordered-logit models with one observation per participant and robust standard errors. Citizens' behavioral measures remain unrelated to perceived corruption. Among judges, the signs of all coefficients are preserved: punishment inflicted is statistically significant ($p=0.015$), while bribes accepted ($p=0.085$) and the bribe rejection share ($p=0.073$) are significant at the 10 percent level.

\begingroup
\small
\begin{longtable}{lccc}
\caption{\normalsize Ordered-logit specifications}
\label{tab:ologit}\\
\toprule
& (1) & (2) & (3) \\
& Citizens & Judges & Rejection share \\
\midrule
\endfirsthead

\multicolumn{4}{l}{\small\textit{Table \thetable\ continued from previous page}}\\
\toprule
& (1) & (2) & (3) \\
& Citizens & Judges & Rejection share \\
\midrule
\endhead

\midrule
\multicolumn{4}{r}{\small\textit{Continued on next page}}\\
\endfoot

\bottomrule
\endlastfoot

Bribe offered
& -0.0301 &  &  \\
& (0.1137) &  &  \\

Contribution
& -0.0113 &  &  \\
& (0.0269) &  &  \\

Punishment received
& 0.0050 &  &  \\
& (0.0440) &  &  \\

Bribes accepted
&  & 0.1684$^{*}$ &  \\
&  & (0.0978) &  \\

Punishment inflicted
&  & 0.2615$^{**}$ &  \\
&  & (0.1078) &  \\

Bribe rejection share
&  &  & -1.9727$^{*}$ \\
&  &  & (1.0984) \\

\addlinespace
PENDO
& -0.4816 & -0.0369 & 0.1434 \\
& (0.4233) & (1.0265) & (1.0381) \\

PENDO\_EXIT
& -0.8645$^{**}$ & 0.1288 & -0.1516 \\
& (0.4267) & (0.9884) & (1.0324) \\

PINFO
& -0.3243 & -0.5521 & -0.4705 \\
& (0.4286) & (1.0107) & (1.0520) \\

\addlinespace
Cut point 1
& -1.6670$^{***}$ & 0.5374 & -1.0806 \\
& (0.4815) & (0.9965) & (1.0069) \\

Cut point 2
& 1.9389$^{***}$ & 4.1598$^{***}$ & 2.2795$^{**}$ \\
& (0.4873) & (1.1922) & (1.1250) \\

\midrule
Participants
& 295 & 99 & 99 \\

Pseudo $R^{2}$
& 0.0104 & 0.0722 & 0.0236 \\

\end{longtable}
\endgroup

{\footnotesize\noindent\textit{Notes:} Ordered-logit estimates using one observation per participant. The dependent variable is perceived corruption, coded from 1 to 4, with higher values indicating greater perceived corruption. Column (1) includes the three citizen behavioral measures jointly and column (2) includes bribes accepted and punishment inflicted jointly. Column (3) uses the judge-level bribe rejection share. Behavioral measures are averaged within participant over all periods of the bribery game in which the participant takes part in the interaction. Periods of the Exit condition in which the common project is not implemented because two or more citizens withdraw are treated as missing. All specifications include indicators for experimental condition, and robust standard errors are reported in parentheses. FULL (Full Information) is the omitted experimental condition. PINFO denotes Control, PENDO denotes Transparency, and PENDO\_EXIT denotes Exit. Coefficients are ordered-logit coefficients and are therefore not directly comparable in magnitude with the OLS coefficients reported in the main analysis. $^{*}p<0.10$, $^{**}p<0.05$, $^{***}p<0.01$.\par}

\hspace{1cm}

% ================================================================
% G. PERIOD-LEVEL SPECIFICATIONS
% ================================================================

\section{Period-Level Specifications}
\label{app:period}

The main analysis uses one observation per participant. This section instead retains the period-level realizations of experimental behavior: each period of each participant enters as a separate observation, the participant's corruption perception is assigned to all of that participant's periods, and standard errors are clustered at the participant level. Because the outcome does not vary within participant, clustering makes inference valid, but the two approaches are not numerically equivalent. Within-participant variation in behavior cannot be associated with an outcome that is constant within participant, so period-level coefficients are attenuated relative to participant-level coefficients, approximately in proportion to the share of the variance of each behavioral measure that lies between participants (0.62 for bribes accepted and 0.52 for punishment inflicted). In addition, participants observed for more periods receive more weight.

Table~\ref{tab:period_level_roles} reports the results. Among citizens, the behavioral coefficients remain close to zero and the joint null cannot be rejected ($p=0.997$ and $p=0.974$). Among judges, bribes accepted and punishment inflicted are positively associated with perceived corruption and jointly significant in both specifications ($p=0.015$ and $p=0.007$). As expected, the period-level coefficients are smaller than their participant-level counterparts.

\begingroup
\small
\begin{longtable}{p{6.8cm}cc}
\caption{\normalsize Period-level specifications across roles}
\label{tab:period_level_roles}\\
\toprule
& (1) & (2) \\
& Baseline & Background \\
\midrule
\endfirsthead

\multicolumn{3}{l}{\small\textit{Table \thetable\ continued from previous page}}\\
\toprule
& (1) & (2) \\
& Baseline & Background \\
\midrule
\endhead

\midrule
\multicolumn{3}{r}{\small\textit{Continued on next page}}\\
\endfoot

\bottomrule
\endlastfoot

\multicolumn{3}{l}{\textit{Panel A. Citizens}} \\
\addlinespace

Bribe offered
& 0.0008
& 0.0068 \\
& (0.0166)
& (0.0155) \\

Contribution
& -0.0009
& -0.0008 \\
& (0.0043)
& (0.0040) \\

Punishment received
& -0.0000
& -0.0000 \\
& (0.0050)
& (0.0044) \\

\addlinespace
Joint test of behavioral coefficients, $p$-value
& 0.997
& 0.974 \\

Experimental-condition controls
& Yes
& Yes \\

Background controls
& No
& Yes \\

Observations
& 3,596
& 3,596 \\

Participant clusters
& 295
& 295 \\

$R^{2}$
& 0.0156
& 0.1267 \\

\midrule

\multicolumn{3}{l}{\textit{Panel B. Judges}} \\
\addlinespace

Bribes accepted
& 0.0304$^{**}$
& 0.0332$^{**}$ \\
& (0.0142)
& (0.0129) \\

Punishment inflicted
& 0.0389$^{**}$
& 0.0412$^{***}$ \\
& (0.0153)
& (0.0148) \\

\addlinespace
Joint test of behavioral coefficients, $p$-value
& 0.015
& 0.007 \\

Experimental-condition controls
& Yes
& Yes \\

Background controls
& No
& Yes \\

Observations
& 1,240
& 1,240 \\

Participant clusters
& 99
& 99 \\

$R^{2}$
& 0.0806
& 0.1896 \\

\end{longtable}
\endgroup

{\footnotesize\noindent\textit{Notes:} OLS estimates using participant-period observations. The dependent variable is perceived corruption, coded from 1 to 4, with higher values indicating greater perceived corruption. Each period of each participant enters as a separate observation, with the participant's single corruption-perception response assigned to all of that participant's periods; the sample comprises all periods of the bribery game in which the participant takes part in the interaction. Periods of the Exit condition in which the common project is not implemented because two or more citizens withdraw are treated as missing. Column (1) includes the role-specific behavioral measures jointly together with experimental-condition indicators. Column (2) additionally controls for age, gender, association membership, voting status, and blood-donation status. Coefficients on controls are suppressed. Standard errors clustered at the participant level are reported in parentheses. The joint-test rows report $p$-values from Wald tests that the behavioral coefficients are jointly equal to zero. FULL (Full Information) is the omitted experimental condition. PINFO denotes Control, PENDO denotes Transparency, and PENDO\_EXIT denotes Exit. $^{*}p<0.10$, $^{**}p<0.05$, $^{***}p<0.01$.\par}

\hspace{1cm}

% ================================================================
% H. FIXED-GROUP CLUSTERED INFERENCE
% ================================================================

\section{Fixed-Group Clustered Inference}
\label{app:group_cluster}

The three citizens in each group interact repeatedly with the same judge and share the same group history. As an additional robustness check, Table~\ref{tab:group_cluster_citizens} re-estimates the citizen specifications clustering standard errors at the fixed-group level, which allows for arbitrary dependence across citizens belonging to the same experimental group. The experiment contains 99 fixed groups. The results are substantively unchanged: bribes offered, contributions, and punishment received remain very small and statistically indistinguishable from zero in all specifications, and the joint null is not rejected ($p=0.920$, $p=0.972$, and $p=0.998$).

For judges, each fixed group contains exactly one judge and the main specifications use one observation per judge. Clustering at the fixed-group level is therefore equivalent to the robust standard errors reported in Table~\ref{tab:complete_judges}, and the judge estimates are unchanged by construction.

\begingroup
\small
\begin{longtable}{p{6.8cm}ccc}
\caption{\normalsize Fixed-group clustered specifications: Citizens}
\label{tab:group_cluster_citizens}\\
\toprule
& (1) & (2) & (3) \\
& Baseline & Background & + Attitudinal controls \\
\midrule
\endfirsthead

\multicolumn{4}{l}{\small\textit{Table \thetable\ continued from previous page}}\\
\toprule
& (1) & (2) & (3) \\
& Baseline & Background & + Attitudinal controls \\
\midrule
\endhead

\midrule
\multicolumn{4}{r}{\small\textit{Continued on next page}}\\
\endfoot

\bottomrule
\endlastfoot

Bribe offered
& -0.0101 & 0.0005 & 0.0032 \\
& (0.0281) & (0.0258) & (0.0249) \\

Contribution
& -0.0027 & -0.0034 & -0.0006 \\
& (0.0076) & (0.0075) & (0.0073) \\

Punishment received
& 0.0022 & -0.0002 & 0.0010 \\
& (0.0111) & (0.0099) & (0.0102) \\

\addlinespace
Age
&  & -0.0251$^{**}$ & -0.0209$^{*}$ \\
&  & (0.0107) & (0.0112) \\

Gender: Male
&  & -0.3204$^{***}$ & -0.3454$^{***}$ \\
&  & (0.0637) & (0.0663) \\

Gender: Not answering
&  & -0.1598 & -0.1241 \\
&  & (0.1742) & (0.1752) \\

Association membership: Yes
&  & -0.1820$^{**}$ & -0.1586$^{*}$ \\
&  & (0.0816) & (0.0817) \\

Voting status: No
&  & 0.1163 & 0.0941 \\
&  & (0.0963) & (0.0986) \\

Voting status: I can't
&  & -0.0867 & -0.1003 \\
&  & (0.0987) & (0.1012) \\

Blood-donation status: No
&  & -0.0115 & -0.0239 \\
&  & (0.0834) & (0.0892) \\

Blood-donation status: Did not meet requirements / age
&  & 0.0772 & 0.0448 \\
&  & (0.0912) & (0.0949) \\

\addlinespace
Generalized trust
&  &  & -0.0394 \\
&  &  & (0.0277) \\

Justifiability: Undeserved benefits
&  &  & -0.0379 \\
&  &  & (0.0333) \\

Justifiability: Fare evasion
&  &  & -0.0187 \\
&  &  & (0.0278) \\

Justifiability: Tax evasion
&  &  & 0.0713$^{**}$ \\
&  &  & (0.0276) \\

Justifiability: Accepting a bribe
&  &  & 0.0202 \\
&  &  & (0.0250) \\

\addlinespace
PENDO
& -0.1172 & -0.1386 & -0.1619$^{*}$ \\
& (0.1050) & (0.0894) & (0.0864) \\

PENDO\_EXIT
& -0.2213$^{**}$ & -0.2847$^{***}$ & -0.2885$^{***}$ \\
& (0.1075) & (0.0966) & (0.0906) \\

PINFO
& -0.0773 & -0.1041 & -0.1085 \\
& (0.1089) & (0.0991) & (0.0973) \\

\addlinespace
Constant
& 2.9666$^{***}$ & 3.7729$^{***}$ & 3.7242$^{***}$ \\
& (0.1146) & (0.2762) & (0.2904) \\

\midrule
Participants
& 295 & 295 & 295 \\

Fixed-group clusters
& 99 & 99 & 99 \\

$R^{2}$
& 0.0169 & 0.1233 & 0.1628 \\

\end{longtable}
\endgroup

{\footnotesize\noindent\textit{Notes:} OLS estimates using one observation per citizen. The dependent variable is perceived corruption, coded from 1 to 4, with higher values indicating greater perceived corruption. Behavioral measures are averaged within participant over all periods of the bribery game in which the participant takes part in the interaction. Periods of the Exit condition in which the common project is not implemented because two or more citizens withdraw are treated as missing. Standard errors clustered at the fixed-group level are reported in parentheses, allowing arbitrary dependence across citizens belonging to the same fixed group. Female is the reference category for gender, No for association membership, Yes for voting status, Yes for blood-donation status, and FULL (Full Information) for the experimental condition. PINFO denotes Control, PENDO denotes Transparency, and PENDO\_EXIT denotes Exit. Generalized trust and the four justifiability measures enter on their original six-point scales. $^{*}p<0.10$, $^{**}p<0.05$, $^{***}p<0.01$.\par}

\hspace{1cm}

% ================================================================
% I. SESSION FIXED-EFFECTS SPECIFICATIONS
% ================================================================

\section{Session Fixed-Effects Specifications}
\label{app:session_fe}

The main specifications include indicators for the four experimental conditions. As an additional robustness check, this section replaces them with a complete set of session indicators, which allows each of the 14 experimental sessions to have its own intercept and therefore absorbs both condition-level differences and any other differences across sessions. Since the experimental condition is constant within each session, condition and session indicators cannot be separately identified. Session 1 is the omitted reference category. All other aspects of the main analysis are unchanged.

\subsection{Citizens}
\label{app:session_fe_citizens}

Table~\ref{tab:session_fe_citizens} reports the citizen specifications with session indicators. Bribes offered, contributions, and punishment received remain very small and statistically indistinguishable from zero in all three specifications, and the joint null is not rejected ($p=0.902$, $p=0.952$, and $p=0.994$).

\begingroup
\scriptsize
\begin{longtable}{p{6.5cm}ccc}
\caption{\normalsize Session fixed-effects specifications: Citizens}
\label{tab:session_fe_citizens}\\
\toprule
& (1) & (2) & (3) \\
& Baseline & Background & + Attitudinal controls \\
\midrule
\endfirsthead

\multicolumn{4}{l}{\scriptsize\textit{Table \thetable\ continued from previous page}}\\
\toprule
& (1) & (2) & (3) \\
& Baseline & Background & + Attitudinal controls \\
\midrule
\endhead

\midrule
\multicolumn{4}{r}{\scriptsize\textit{Continued on next page}}\\
\endfoot

\bottomrule
\endlastfoot

Bribe offered
& -0.0121 & -0.0017 & 0.0007 \\
& (0.0298) & (0.0279) & (0.0276) \\

Contribution
& -0.0020 & -0.0038 & -0.0010 \\
& (0.0077) & (0.0074) & (0.0073) \\

Punishment received
& 0.0039 & 0.0007 & 0.0025 \\
& (0.0119) & (0.0107) & (0.0109) \\

\addlinespace
Age
&  & -0.0240$^{**}$ & -0.0205$^{*}$ \\
&  & (0.0108) & (0.0113) \\

Gender: Male
&  & -0.3267$^{***}$ & -0.3448$^{***}$ \\
&  & (0.0702) & (0.0710) \\

Gender: Not answering
&  & -0.1289 & -0.1143 \\
&  & (0.1756) & (0.1739) \\

Association membership: Yes
&  & -0.1979$^{***}$ & -0.1725$^{**}$ \\
&  & (0.0754) & (0.0766) \\

Voting status: No
&  & 0.1098 & 0.0877 \\
&  & (0.1035) & (0.1027) \\

Voting status: I can't
&  & -0.0854 & -0.0995 \\
&  & (0.0976) & (0.1011) \\

Blood-donation status: No
&  & 0.0228 & 0.0060 \\
&  & (0.0895) & (0.0939) \\

Blood-donation status: Did not meet requirements / age
&  & 0.0965 & 0.0662 \\
&  & (0.1027) & (0.1067) \\

\addlinespace
Generalized trust
&  &  & -0.0376 \\
&  &  & (0.0276) \\

Justifiability: Undeserved benefits
&  &  & -0.0383 \\
&  &  & (0.0350) \\

Justifiability: Fare evasion
&  &  & -0.0099 \\
&  &  & (0.0262) \\

Justifiability: Tax evasion
&  &  & 0.0651$^{**}$ \\
&  &  & (0.0268) \\

Justifiability: Accepting a bribe
&  &  & 0.0144 \\
&  &  & (0.0241) \\

\addlinespace
Session 2
& -0.1602 & -0.1643 & -0.1663 \\
& (0.1839) & (0.1804) & (0.1864) \\

Session 3
& -0.1259 & -0.1380 & -0.1090 \\
& (0.1587) & (0.1603) & (0.1646) \\

Session 4
& -0.0166 & -0.0062 & -0.0516 \\
& (0.1735) & (0.1620) & (0.1636) \\

Session 5
& -0.2053 & -0.1739 & -0.1802 \\
& (0.1691) & (0.1485) & (0.1487) \\

Session 6
& -0.0806 & -0.0977 & -0.0708 \\
& (0.1544) & (0.1554) & (0.1576) \\

Session 7
& 0.0907 & 0.1182 & 0.1059 \\
& (0.1884) & (0.1755) & (0.1711) \\

Session 8
& -0.2185 & -0.2261 & -0.2086 \\
& (0.1640) & (0.1529) & (0.1581) \\

Session 9
& -0.0378 & -0.0328 & 0.0077 \\
& (0.1636) & (0.1448) & (0.1519) \\

Session 10
& -0.1048 & -0.1840 & -0.1695 \\
& (0.1584) & (0.1544) & (0.1581) \\

Session 11
& -0.3700$^{**}$ & -0.4325$^{**}$ & -0.3824$^{**}$ \\
& (0.1724) & (0.1713) & (0.1719) \\

Session 12
& 0.0887 & 0.1562 & 0.1200 \\
& (0.2025) & (0.1888) & (0.1734) \\

Session 13
& -0.2229 & -0.2362 & -0.2225 \\
& (0.1767) & (0.1716) & (0.1795) \\

Session 14
& -0.2082 & -0.2230 & -0.2396 \\
& (0.1736) & (0.1761) & (0.1752) \\

\addlinespace
Constant
& 2.9562$^{***}$ & 3.7127$^{***}$ & 3.6596$^{***}$ \\
& (0.1518) & (0.2885) & (0.3082) \\

\midrule
Participants
& 295 & 295 & 295 \\

$R^{2}$
& 0.0448 & 0.1537 & 0.1842 \\

\end{longtable}
\endgroup

{\footnotesize\noindent\textit{Notes:} OLS estimates using one observation per citizen. The dependent variable is perceived corruption, coded from 1 to 4, with higher values indicating greater perceived corruption. Behavioral measures are averaged within participant over all periods of the bribery game in which the participant takes part in the interaction. Periods of the Exit condition in which the common project is not implemented because two or more citizens withdraw are treated as missing. All specifications include a complete set of session indicators in place of the experimental-condition indicators used in the main analysis; Session 1 is the omitted reference category. Column (2) additionally controls for age, gender, association membership, voting status, and blood-donation status. Column (3) further controls for generalized trust and the stated justifiability of claiming undeserved benefits, fare evasion, tax evasion, and accepting a bribe. Robust standard errors are reported in parentheses. $^{*}p<0.10$, $^{**}p<0.05$, $^{***}p<0.01$.\par}

\hspace{1cm}

\subsection{Judges}
\label{app:session_fe_judges}

Table~\ref{tab:session_fe_judges} reports the corresponding specifications for judges. Punishment inflicted remains positive and statistically significant throughout ($p=0.003$, $p=0.005$, and $p=0.010$). The coefficient on bribes accepted remains positive and similar in magnitude to the main specifications but is estimated less precisely ($p=0.121$, $p=0.062$, and $p=0.084$). The two measures remain jointly significant ($p=0.005$, $p=0.006$, and $p=0.013$).

\begingroup
\scriptsize
\begin{longtable}{p{6.5cm}ccc}
\caption{\normalsize Session fixed-effects specifications: Judges}
\label{tab:session_fe_judges}\\
\toprule
& (1) & (2) & (3) \\
& Baseline & Background & + Attitudinal controls \\
\midrule
\endfirsthead

\multicolumn{4}{l}{\scriptsize\textit{Table \thetable\ continued from previous page}}\\
\toprule
& (1) & (2) & (3) \\
& Baseline & Background & + Attitudinal controls \\
\midrule
\endhead

\midrule
\multicolumn{4}{r}{\scriptsize\textit{Continued on next page}}\\
\endfoot

\bottomrule
\endlastfoot

Bribes accepted
& 0.0391 & 0.0477$^{*}$ & 0.0452$^{*}$ \\
& (0.0250) & (0.0251) & (0.0258) \\

Punishment inflicted
& 0.0794$^{***}$ & 0.0837$^{***}$ & 0.0801$^{**}$ \\
& (0.0256) & (0.0289) & (0.0304) \\

\addlinespace
Age
&  & -0.0222$^{*}$ & -0.0185 \\
&  & (0.0113) & (0.0124) \\

Gender: Male
&  & -0.2624$^{*}$ & -0.2919$^{**}$ \\
&  & (0.1522) & (0.1430) \\

Gender: Not answering
&  & 0.3715 & 0.3654 \\
&  & (0.2714) & (0.3314) \\

Gender: Not identified
&  & -0.0129 & 0.0047 \\
&  & (0.1595) & (0.1789) \\

Association membership: Yes
&  & -0.2108 & -0.2114 \\
&  & (0.1407) & (0.1377) \\

Voting status: No
&  & -0.0050 & -0.0334 \\
&  & (0.1639) & (0.1637) \\

Voting status: I can't
&  & 0.1460 & 0.1656 \\
&  & (0.2827) & (0.2752) \\

Blood-donation status: No
&  & 0.1134 & 0.0797 \\
&  & (0.1406) & (0.1414) \\

Blood-donation status: Did not meet requirements / age
&  & 0.1956 & 0.1664 \\
&  & (0.1924) & (0.2107) \\

\addlinespace
Generalized trust
&  &  & -0.0544 \\
&  &  & (0.0711) \\

Justifiability: Undeserved benefits
&  &  & -0.0198 \\
&  &  & (0.0677) \\

Justifiability: Fare evasion
&  &  & 0.0189 \\
&  &  & (0.0570) \\

Justifiability: Tax evasion
&  &  & 0.0274 \\
&  &  & (0.0558) \\

Justifiability: Accepting a bribe
&  &  & 0.0249 \\
&  &  & (0.0609) \\

\addlinespace
Session 2
& -0.1523 & -0.1594 & -0.1756 \\
& (0.2469) & (0.2069) & (0.2153) \\

Session 3
& -0.2170 & -0.2370 & -0.2172 \\
& (0.2518) & (0.2583) & (0.2682) \\

Session 4
& -0.2033 & -0.2423 & -0.2138 \\
& (0.2590) & (0.2772) & (0.2645) \\

Session 5
& 0.2425 & 0.2036 & 0.2570 \\
& (0.2531) & (0.2598) & (0.2753) \\

Session 6
& 0.1604 & 0.0341 & 0.0090 \\
& (0.2940) & (0.3191) & (0.2999) \\

Session 7
& -0.3328 & -0.2901 & -0.3217 \\
& (0.3123) & (0.3189) & (0.3249) \\

Session 8
& -0.0340 & -0.1103 & -0.0732 \\
& (0.2617) & (0.2508) & (0.2932) \\

Session 9
& 0.1269 & 0.2568 & 0.2284 \\
& (0.3834) & (0.3720) & (0.4008) \\

Session 10
& -0.1270 & -0.1212 & -0.1375 \\
& (0.2488) & (0.2406) & (0.2414) \\

Session 11
& 0.0933 & -0.0837 & -0.0916 \\
& (0.2398) & (0.2467) & (0.2604) \\

Session 12
& -0.1272 & -0.1348 & -0.0462 \\
& (0.3122) & (0.2410) & (0.2580) \\

Session 13
& 0.0178 & 0.0450 & 0.0166 \\
& (0.2506) & (0.2231) & (0.2492) \\

Session 14
& 0.2780 & 0.1513 & 0.0831 \\
& (0.3506) & (0.3987) & (0.4533) \\

\addlinespace
Constant
& 2.3685$^{***}$ & 2.9876$^{***}$ & 3.0129$^{***}$ \\
& (0.2443) & (0.3917) & (0.5268) \\

\midrule
Participants
& 99 & 99 & 99 \\

$R^{2}$
& 0.1893 & 0.2863 & 0.3015 \\

\end{longtable}
\endgroup

{\footnotesize\noindent\textit{Notes:} OLS estimates using one observation per judge. The dependent variable is perceived corruption, coded from 1 to 4, with higher values indicating greater perceived corruption. Behavioral measures are averaged within participant over all periods of the bribery game in which the participant takes part in the interaction. Periods of the Exit condition in which the common project is not implemented because two or more citizens withdraw are treated as missing. All specifications include a complete set of session indicators in place of the experimental-condition indicators used in the main analysis; Session 1 is the omitted reference category. Column (2) additionally controls for age, gender, association membership, voting status, and blood-donation status. Column (3) further controls for generalized trust and the stated justifiability of claiming undeserved benefits, fare evasion, tax evasion, and accepting a bribe. Robust standard errors are reported in parentheses. $^{*}p<0.10$, $^{**}p<0.05$, $^{***}p<0.01$.\par}

\hspace{1cm}

\subsection{Judges' Revealed Resistance to Bribery}
\label{app:session_fe_rejection}

Table~\ref{tab:session_fe_rejection} replaces the experimental-condition indicators with session indicators in the rejection-share specifications. The estimated association remains negative and similar in magnitude, although statistical precision decreases: the coefficient is $-0.577$ in the baseline specification ($p=0.087$), $-0.552$ after adding background characteristics ($p=0.101$), and $-0.461$ in the specification with additional attitudinal controls ($p=0.194$).

\begingroup
\scriptsize
\begin{longtable}{p{6.5cm}ccc}
\caption{\normalsize Session fixed-effects specifications: Judges' revealed resistance to bribery}
\label{tab:session_fe_rejection}\\
\toprule
& (1) & (2) & (3) \\
& Baseline & Background & + Attitudinal controls \\
\midrule
\endfirsthead

\multicolumn{4}{l}{\scriptsize\textit{Table \thetable\ continued from previous page}}\\
\toprule
& (1) & (2) & (3) \\
& Baseline & Background & + Attitudinal controls \\
\midrule
\endhead

\midrule
\multicolumn{4}{r}{\scriptsize\textit{Continued on next page}}\\
\endfoot

\bottomrule
\endlastfoot

Bribe rejection share
& -0.5768$^{*}$ & -0.5516 & -0.4606 \\
& (0.3328) & (0.3319) & (0.3508) \\

\addlinespace
Age
&  & -0.0189 & -0.0148 \\
&  & (0.0130) & (0.0140) \\

Gender: Male
&  & -0.2848$^{*}$ & -0.3151$^{**}$ \\
&  & (0.1524) & (0.1469) \\

Gender: Not answering
&  & 0.2090 & 0.2201 \\
&  & (0.2854) & (0.3463) \\

Gender: Not identified
&  & 0.2015 & 0.2241 \\
&  & (0.1728) & (0.1884) \\

Association membership: Yes
&  & -0.0898 & -0.0971 \\
&  & (0.1585) & (0.1565) \\

Voting status: No
&  & 0.0325 & -0.0047 \\
&  & (0.1963) & (0.1976) \\

Voting status: I can't
&  & 0.1808 & 0.2089 \\
&  & (0.2876) & (0.2781) \\

Blood-donation status: No
&  & 0.0931 & 0.0552 \\
&  & (0.1612) & (0.1651) \\

Blood-donation status: Did not meet requirements / age
&  & 0.0542 & 0.0367 \\
&  & (0.2288) & (0.2450) \\

\addlinespace
Generalized trust
&  &  & -0.0667 \\
&  &  & (0.0739) \\

Justifiability: Undeserved benefits
&  &  & -0.0050 \\
&  &  & (0.0791) \\

Justifiability: Fare evasion
&  &  & 0.0078 \\
&  &  & (0.0597) \\

Justifiability: Tax evasion
&  &  & 0.0356 \\
&  &  & (0.0609) \\

Justifiability: Accepting a bribe
&  &  & 0.0304 \\
&  &  & (0.0597) \\

\addlinespace
Session 2
& -0.0938 & -0.0856 & -0.0985 \\
& (0.2492) & (0.2212) & (0.2415) \\

Session 3
& -0.2864 & -0.2760 & -0.2559 \\
& (0.2513) & (0.2524) & (0.2814) \\

Session 4
& -0.0848 & -0.0721 & -0.0456 \\
& (0.2665) & (0.2584) & (0.2601) \\

Session 5
& 0.2491 & 0.2518 & 0.3134 \\
& (0.2505) & (0.2665) & (0.2997) \\

Session 6
& 0.0759 & -0.0221 & -0.0439 \\
& (0.2798) & (0.3045) & (0.3026) \\

Session 7
& -0.3090 & -0.2177 & -0.2580 \\
& (0.2877) & (0.3181) & (0.3234) \\

Session 8
& 0.0516 & 0.0818 & 0.1018 \\
& (0.3462) & (0.3309) & (0.3614) \\

Session 9
& 0.0600 & 0.2286 & 0.2071 \\
& (0.4107) & (0.4198) & (0.4344) \\

Session 10
& -0.2239 & -0.1715 & -0.1902 \\
& (0.2543) & (0.2518) & (0.2628) \\

Session 11
& 0.0070 & -0.1117 & -0.1178 \\
& (0.2424) & (0.2633) & (0.2913) \\

Session 12
& -0.0001 & 0.0366 & 0.1197 \\
& (0.2295) & (0.2598) & (0.2811) \\

Session 13
& -0.0441 & -0.0203 & -0.0586 \\
& (0.2423) & (0.2061) & (0.2596) \\

Session 14
& 0.1066 & 0.0133 & -0.0538 \\
& (0.3219) & (0.3663) & (0.4278) \\

\addlinespace
Constant
& 2.8354$^{***}$ & 3.3342$^{***}$ & 3.3448$^{***}$ \\
& (0.1776) & (0.4249) & (0.5414) \\

\midrule
Participants
& 99 & 99 & 99 \\

$R^{2}$
& 0.0994 & 0.1839 & 0.2050 \\

\end{longtable}
\endgroup

{\footnotesize\noindent\textit{Notes:} OLS estimates using one observation per judge. The dependent variable is perceived corruption, coded from 1 to 4, with higher values indicating greater perceived corruption. Bribe rejection share is the proportion of positive bribe offers rejected by each judge over all periods of the bribery game. All specifications include a complete set of session indicators in place of the experimental-condition indicators used in the main analysis; Session 1 is the omitted reference category. Column (2) additionally controls for age, gender, association membership, voting status, and blood-donation status. Column (3) further controls for generalized trust and the stated justifiability of claiming undeserved benefits, fare evasion, tax evasion, and accepting a bribe. Robust standard errors are reported in parentheses. $^{*}p<0.10$, $^{**}p<0.05$, $^{***}p<0.01$.\par}

\hspace{1cm}

% ================================================================
% J. CONTROLLING FOR BRIBE OPPORTUNITIES RECEIVED BY JUDGES
% ================================================================

\section{Controlling for Bribe Opportunities Received by Judges}
\label{app:bribe_opportunities}

The amount of bribery accepted by a judge depends not only on the judge's decision to accept or reject bribes but also on the bribery opportunities generated by citizens. Table~\ref{tab:bribe_opportunities} therefore augments the main judge specifications with the total amount of bribes offered to the judge. Since the amount accepted equals the amount offered minus the amount rejected, this specification is algebraically equivalent to one that includes the amount offered and the amount rejected: the coefficient on bribes accepted reported in the table is the coefficient on the amount rejected with the opposite sign, and the coefficient on the amount offered in that alternative form is the sum of the two bribery coefficients reported in the table.

The coefficient on bribes accepted is 0.1082 in the baseline specification ($p=0.002$), 0.1030 with background controls ($p=0.002$), and 0.0936 with additional attitudinal controls ($p=0.009$). In the equivalent form, the coefficient on the amount rejected is $-0.1082$ in the baseline specification, while the coefficient on the amount offered is 0.0312 ($p=0.218$). Punishment inflicted remains positive and statistically significant throughout. This table and the rejection-share specifications of Table~\ref{tab:complete_rejection} therefore describe the same margin of judge behavior and should not be read as independent pieces of evidence. \begingroup
\small
\begin{longtable}{p{6.8cm}ccc}
\caption{\normalsize Judges: Controlling for Bribe Opportunities Received}
\label{tab:bribe_opportunities}\\
\toprule
& (1) & (2) & (3) \\
& Baseline & Background & + Attitudinal controls \\
\midrule
\endfirsthead

\multicolumn{4}{l}{\small\textit{Table \thetable\ continued from previous page}}\\
\toprule
& (1) & (2) & (3) \\
& Baseline & Background & + Attitudinal controls \\
\midrule
\endhead

\midrule
\multicolumn{4}{r}{\small\textit{Continued on next page}}\\
\endfoot

\bottomrule
\endlastfoot

Bribes accepted
& 0.1082$^{***}$ & 0.1030$^{***}$ & 0.0936$^{***}$ \\
& (0.0336) & (0.0324) & (0.0347) \\

Total bribes offered
& -0.0770$^{**}$ & -0.0609$^{*}$ & -0.0526 \\
& (0.0349) & (0.0360) & (0.0372) \\

Punishment inflicted
& 0.0770$^{***}$ & 0.0788$^{***}$ & 0.0776$^{***}$ \\
& (0.0251) & (0.0263) & (0.0278) \\

\addlinespace
Age
&  & -0.0184$^{*}$ & -0.0155 \\
&  & (0.0108) & (0.0117) \\

Gender: Male
&  & -0.2862$^{**}$ & -0.3056$^{**}$ \\
&  & (0.1310) & (0.1269) \\

Gender: Not answering
&  & 0.2940$^{*}$ & 0.2814 \\
&  & (0.1649) & (0.2233) \\

Gender: Not identified
&  & 0.0676 & 0.0683 \\
&  & (0.1836) & (0.2038) \\

Association membership: Yes
&  & -0.2373$^{*}$ & -0.2397$^{*}$ \\
&  & (0.1390) & (0.1343) \\

Voting status: No
&  & -0.1354 & -0.1567 \\
&  & (0.1623) & (0.1616) \\

Voting status: I can't
&  & 0.1740 & 0.1936 \\
&  & (0.2362) & (0.2417) \\

Blood-donation status: No
&  & 0.1335 & 0.1047 \\
&  & (0.1400) & (0.1393) \\

Blood-donation status: Did not meet requirements / age
&  & 0.1521 & 0.1350 \\
&  & (0.1770) & (0.1847) \\

\addlinespace
Generalized trust
&  &  & -0.0377 \\
&  &  & (0.0661) \\

Justifiability: Undeserved benefits
&  &  & -0.0025 \\
&  &  & (0.0637) \\

Justifiability: Fare evasion
&  &  & 0.0212 \\
&  &  & (0.0536) \\

Justifiability: Tax evasion
&  &  & 0.0113 \\
&  &  & (0.0495) \\

Justifiability: Accepting a bribe
&  &  & 0.0225 \\
&  &  & (0.0590) \\

\addlinespace
PENDO
& -0.0735 & -0.1899 & -0.1799 \\
& (0.2464) & (0.2579) & (0.2698) \\

PENDO\_EXIT
& -0.0710 & -0.1885 & -0.1978 \\
& (0.2507) & (0.2694) & (0.3197) \\

PINFO
& -0.2271 & -0.3118 & -0.3013 \\
& (0.2460) & (0.2633) & (0.2770) \\

\addlinespace
Constant
& 2.5611$^{***}$ & 3.1768$^{***}$ & 3.1354$^{***}$ \\
& (0.2707) & (0.4812) & (0.5943) \\

\midrule
Participants
& 99 & 99 & 99 \\

$R^{2}$
& 0.1471 & 0.2522 & 0.2638 \\

\end{longtable}
\endgroup

{\footnotesize\noindent\textit{Notes:} OLS estimates using one observation per judge. The dependent variable is perceived corruption, coded from 1 to 4, with higher values indicating greater perceived corruption. Behavioral measures are averaged within participant over all periods of the bribery game in which the participant takes part in the interaction. Periods of the Exit condition in which the common project is not implemented because two or more citizens withdraw are treated as missing. Bribes accepted is the average amount of bribery accepted by the judge per period. Total bribes offered is the average of the sum of the bribe offers made by the citizens to the judge per period. Because accepted bribes equal offered bribes minus rejected bribes, replacing bribes accepted with the amount rejected yields the same coefficient with the opposite sign, and a coefficient on total bribes offered equal to the sum of the two coefficients reported here. Robust standard errors are reported in parentheses. Female is the reference category for gender, No for association membership, Yes for voting status, Yes for blood-donation status, and FULL (Full Information) for the experimental condition. PINFO denotes Control, PENDO denotes Transparency, and PENDO\_EXIT denotes Exit. Generalized trust and the four justifiability measures enter on their original six-point scales. $^{*}p<0.10$, $^{**}p<0.05$, $^{***}p<0.01$.\par}

\hspace{1cm}

Table~\ref{tab:judge_sensitivity} reports this equivalent form together with four additional checks. Column (1) shows that the amount of bribery offered to the judge, entered alone, is not associated with perceived corruption ($p=0.354$). Column (2) is the equivalent form of the baseline specification of Table~\ref{tab:bribe_opportunities}. The distribution of the amount rejected is strongly right-skewed: its median is 0.09 points per period, its 95th percentile 3.42, and its maximum 9.80. When the three judges with the largest amounts rejected are excluded in column (3), the coefficient on the amount rejected falls to $-0.056$ and is no longer statistically significant ($p=0.386$), while punishment inflicted remains significant ($p=0.011$). In the rank-based specification of column (4), the amount rejected remains negatively associated with perceived corruption ($p=0.019$), as does punishment inflicted positively ($p=0.006$). Column (5) restricts the sample to the 52 judges who reject at least one positive offer: among them, the share of offers rejected is unrelated to perceived corruption ($p=0.594$), so that the rejection-share result reflects mainly the difference between judges who never reject an offer and judges who do.

\begingroup
\small
\begin{longtable}{p{4.3cm}ccccc}
\caption{\normalsize Judges: bribery opportunities, rejection, and sensitivity checks}
\label{tab:judge_sensitivity}\\
\toprule
& (1) & (2) & (3) & (4) & (5) \\
& Offered & Offered and & Excluding & Ranks & Judges who \\
& only & rejected & three judges & & reject \\
\midrule
\endfirsthead

\multicolumn{6}{l}{\small\textit{Table \thetable\ continued from previous page}}\\
\toprule
& (1) & (2) & (3) & (4) & (5) \\
& Offered & Offered and & Excluding & Ranks & Judges who \\
& only & rejected & three judges & & reject \\
\midrule
\endhead

\midrule
\multicolumn{6}{r}{\small\textit{Continued on next page}}\\
\endfoot

\bottomrule
\endlastfoot

Total bribes offered
& 0.0228 & 0.0312 & 0.0306 & 0.1062 &  \\
& (0.0244) & (0.0251) & (0.0256) & (0.1005) &  \\

Amount rejected
&  & -0.1082$^{***}$ & -0.0559 & -0.2013$^{**}$ &  \\
&  & (0.0336) & (0.0641) & (0.0841) &  \\

Punishment inflicted
&  & 0.0770$^{***}$ & 0.0880$^{**}$ & 0.2294$^{***}$ &  \\
&  & (0.0251) & (0.0338) & (0.0812) &  \\

Bribe rejection share
&  &  &  &  & -0.1942 \\
&  &  &  &  & (0.3615) \\
\midrule
Experimental-condition controls
& Yes & Yes & Yes & Yes & Yes \\
Judges
& 99 & 99 & 96 & 99 & 52 \\
$R^{2}$
& 0.0204 & 0.1471 & 0.1249 & 0.1419 & 0.0544 \\

\end{longtable}
\endgroup

{\footnotesize\noindent\textit{Notes:} OLS estimates using one observation per judge, with robust standard errors in parentheses. The dependent variable is perceived corruption, coded from 1 to 4, with higher values indicating greater perceived corruption. Behavioral measures are averaged within participant over all periods of the bribery game in which the participant takes part in the interaction. Periods of the Exit condition in which the common project is not implemented because two or more citizens withdraw are treated as missing. Column (1) includes the average total amount of bribes offered to the judge as the only behavioral measure. Column (2) is the specification of column (1) of Table~\ref{tab:bribe_opportunities} written in terms of the amount offered and the amount rejected, which equals the amount offered minus the amount accepted. Column (3) excludes the three judges with the largest average amount rejected. In column (4) perceived corruption and the three behavioral measures are replaced by their ranks across judges. Column (5) regresses perceived corruption on the bribe rejection share among the judges who reject at least one positive offer. All specifications include experimental-condition indicators. $^{*}p<0.10$, $^{**}p<0.05$, $^{***}p<0.01$.\par}

\hspace{1cm}

% ================================================================
% K. DIFFERENCE BETWEEN ROLES
% ================================================================

\section{Difference between Roles: Pooled Specifications}
\label{app:role_difference}

The main text reports separate regressions for citizens and judges. A significant association in one role and an insignificant one in the other do not by themselves establish that the two associations differ. This section therefore reports pooled specifications that estimate the difference directly. Citizens and the judge of the same group are not independent observations, so standard errors are clustered at the fixed-group level.

Table~\ref{tab:role_difference} reports three versions of the comparison. In Panel A, each participant's own behavior is standardized within role and all regressors are interacted with a judge indicator. For citizens, the association of perceived corruption with a one-standard-deviation difference in behavior is close to zero for both bribery and punishment; the 95 percent confidence intervals in the baseline specification are $[-0.081,\ 0.048]$ and $[-0.054,\ 0.071]$, so that associations as large as those estimated for judges can be excluded. The difference between roles is $0.150$ for punishment ($p=0.036$) and $0.132$ for bribery ($p=0.083$), and the null hypothesis that both differences are zero is rejected ($p=0.029$, and $p=0.018$ with background controls).

Panel B reports the same differences in original units, that is, per point of bribery and per deduction point. In these units the two differences are not jointly significant at the 5 percent level ($p=0.116$ and $p=0.073$). The original units are not directly comparable across roles: a citizen's offer ranges from 0 to 5 points, whereas the amount accepted by a judge sums over up to three citizens and ranges from 0 to 15, and the standard deviation of the participant-level measure is 1.26 for citizens and 2.80 for judges. For this reason the standardized comparison is used in the main text.

Panel C uses a regressor that is common to the four members of a group, namely a measure of the behavior of the group's judge, and tests whether its association with perceived corruption differs between the judge and the citizens of the same group. The difference in slopes is positive for bribes accepted ($p=0.258$) and for punishment inflicted ($p=0.085$), and negative for the rejection share ($p=0.336$), but none of these differences is statistically significant at the 5 percent level. Taken together, the evidence that the correspondence between perceived corruption and own behavior is stronger for judges than for citizens is moderate, and it is driven mainly by punishment.

\begingroup
\small
\begin{longtable}{p{8.2cm}cc}
\caption{\normalsize Difference between roles: pooled specifications}
\label{tab:role_difference}\\
\toprule
& (1) & (2) \\
& Baseline & Background \\
\midrule
\endfirsthead

\multicolumn{3}{l}{\small\textit{Table \thetable\ continued from previous page}}\\
\toprule
& (1) & (2) \\
& Baseline & Background \\
\midrule
\endhead

\midrule
\multicolumn{3}{r}{\small\textit{Continued on next page}}\\
\endfoot

\bottomrule
\endlastfoot

\multicolumn{3}{l}{\textit{Panel A. Own behavior, standardized within role}} \\
\addlinespace

Bribery: citizens
& -0.0165 & -0.0040 \\
& (0.0327) & (0.0312) \\

Bribery: judges
& 0.1153$^{*}$ & 0.1410$^{**}$ \\
& (0.0662) & (0.0635) \\

Bribery: difference (judges $-$ citizens)
& 0.1318$^{*}$ & 0.1449$^{*}$ \\
& (0.0753) & (0.0745) \\

\addlinespace
Punishment: citizens
& 0.0086 & 0.0020 \\
& (0.0313) & (0.0282) \\

Punishment: judges
& 0.1584$^{**}$ & 0.1680$^{**}$ \\
& (0.0621) & (0.0658) \\

Punishment: difference (judges $-$ citizens)
& 0.1498$^{**}$ & 0.1660$^{**}$ \\
& (0.0705) & (0.0702) \\

\addlinespace
Joint test of the two differences, $p$-value
& 0.029 & 0.018 \\

\midrule

\multicolumn{3}{l}{\textit{Panel B. Own behavior, original units}} \\
\addlinespace

Bribery, per point: difference
& 0.0542 & 0.0535 \\
& (0.0360) & (0.0358) \\

Punishment, per deduction point: difference
& 0.0598 & 0.0708$^{*}$ \\
& (0.0428) & (0.0397) \\

Joint test of the two differences, $p$-value
& 0.116 & 0.073 \\

\midrule

\multicolumn{3}{l}{\textit{Panel C. Common group-level measure: difference in slopes (judges $-$ citizens)}} \\
\addlinespace

Bribes accepted in the group
& 0.0290 & 0.0306 \\
& (0.0254) & (0.0243) \\

Punishment inflicted in the group
& 0.0603$^{*}$ & 0.0585$^{*}$ \\
& (0.0346) & (0.0334) \\

Bribe rejection share in the group
& -0.3831 & -0.3970 \\
& (0.3962) & (0.2965) \\

\midrule
Experimental-condition controls
& Yes & Yes \\
Background controls
& No & Yes \\
Participants
& 394 & 394 \\
Fixed-group clusters
& 99 & 99 \\

\end{longtable}
\endgroup

{\footnotesize\noindent\textit{Notes:} OLS estimates from pooled regressions using one observation per participant, with standard errors clustered at the fixed-group level in parentheses. The dependent variable is perceived corruption, coded from 1 to 4, with higher values indicating greater perceived corruption. Behavioral measures are averaged within participant over all periods of the bribery game in which the participant takes part in the interaction. Periods of the Exit condition in which the common project is not implemented because two or more citizens withdraw are treated as missing. In Panels A and B, bribery is the amount offered for citizens and the amount accepted for judges, and punishment is the number of deduction points received for citizens and inflicted for judges; these measures, the experimental-condition indicators and, in column (2), the background controls are all interacted with a judge indicator, so that the role-specific coefficients coincide with those of separate regressions by role. In Panel A each measure is standardized within role, and coefficients measure the change in perceived corruption associated with a one-standard-deviation change in behavior. Panel B reports the same differences in original units. In Panel C the regressor is a measure of the judge's behavior in the participant's group, common to the four group members, interacted with a judge indicator; controls enter without interactions. Column (2) additionally controls for age, gender, association membership, voting status, and blood-donation status. $^{*}p<0.10$, $^{**}p<0.05$, $^{***}p<0.01$.\par}

\hspace{1cm}

% ================================================================
% L. ALTERNATIVE DEFINITIONS OF THE ESTIMATION SAMPLE
% ================================================================

\section{Alternative Definitions of the Estimation Sample}
\label{app:sample_definition}

The main analysis averages behavior over all periods of the bribery game and treats periods without interaction as missing. Table~\ref{tab:sample_definition} examines two alternative choices. The first excludes the first period, in which the information-purchase option of the Transparency and Exit conditions is not yet available. The second retains, for judges, the 55 periods of the Exit condition in which two or more citizens withdraw, recording zero bribes accepted and zero punishment. Column (4) applies both changes.

The citizen results are unaffected: the joint null is never rejected. The judge behavioral measures remain jointly significant under all definitions (from $p=0.011$ to $p=0.026$), and the joint test of the difference between roles ranges from $p=0.029$ to $p=0.049$. The bribe rejection share is the measure most sensitive to these choices: when the first period is excluded, its coefficient falls from $-0.616$ ($p=0.050$) to $-0.500$ ($p=0.098$).

\begingroup
\small
\begin{longtable}{p{5.6cm}cccc}
\caption{\normalsize Alternative definitions of the estimation sample}
\label{tab:sample_definition}\\
\toprule
& (1) & (2) & (3) & (4) \\
& Main & From second & Non-interaction & Both \\
& & period & periods included & changes \\
\midrule
\endfirsthead

\multicolumn{5}{l}{\small\textit{Table \thetable\ continued from previous page}}\\
\toprule
& (1) & (2) & (3) & (4) \\
& Main & From second & Non-interaction & Both \\
& & period & periods included & changes \\
\midrule
\endhead

\midrule
\multicolumn{5}{r}{\small\textit{Continued on next page}}\\
\endfoot

\bottomrule
\endlastfoot

\multicolumn{5}{l}{\textit{Panel A. Citizens}} \\
\addlinespace

Bribe offered
& -0.0101 & -0.0005 & -0.0101 & -0.0005 \\
& (0.0292) & (0.0276) & (0.0292) & (0.0276) \\

Contribution
& -0.0027 & -0.0022 & -0.0027 & -0.0022 \\
& (0.0071) & (0.0068) & (0.0071) & (0.0068) \\

Punishment received
& 0.0022 & 0.0027 & 0.0022 & 0.0027 \\
& (0.0114) & (0.0108) & (0.0114) & (0.0108) \\
\addlinespace
Joint test, $p$-value
& 0.907 & 0.969 & 0.907 & 0.969 \\
Participants
& 295 & 294 & 295 & 294 \\
\midrule

\multicolumn{5}{l}{\textit{Panel B. Judges}} \\
\addlinespace

Bribes accepted
& 0.0412$^{*}$ & 0.0417$^{*}$ & 0.0369 & 0.0368 \\
& (0.0239) & (0.0227) & (0.0238) & (0.0228) \\

Punishment inflicted
& 0.0687$^{**}$ & 0.0666$^{***}$ & 0.0658$^{**}$ & 0.0650$^{**}$ \\
& (0.0273) & (0.0252) & (0.0283) & (0.0277) \\
\addlinespace
Joint test, $p$-value
& 0.016 & 0.011 & 0.026 & 0.021 \\
Participants
& 99 & 99 & 99 & 99 \\
\midrule

\multicolumn{5}{l}{\textit{Panel C. Judges' bribe rejection share}} \\
\addlinespace

Bribe rejection share
& -0.6156$^{*}$ & -0.5001$^{*}$ & -0.6156$^{*}$ & -0.5001$^{*}$ \\
& (0.3100) & (0.2996) & (0.3100) & (0.2996) \\
\midrule

\multicolumn{5}{l}{\textit{Panel D. Difference between roles, standardized behavior}} \\
\addlinespace

Joint test of the two differences, $p$-value
& 0.029 & 0.029 & 0.043 & 0.049 \\

\end{longtable}
\endgroup

{\footnotesize\noindent\textit{Notes:} OLS estimates using one observation per participant; all specifications include experimental-condition indicators. The dependent variable is perceived corruption, coded from 1 to 4, with higher values indicating greater perceived corruption. Column (1) reproduces the main specifications. Column (2) averages behavior over all periods after the first. Column (3) retains, for judges, the periods of the Exit condition in which the common project is not implemented because two or more citizens withdraw, recording zero bribes accepted and zero punishment in those periods. Column (4) applies both changes and corresponds to the sample definition used in earlier versions of the paper. The bribe rejection share is computed over the same periods as the other measures in each column; it is unaffected by the treatment of non-interaction periods because no offer is accepted or rejected in them. Robust standard errors are reported in parentheses in Panels A to C; Panel D reports the $p$-value of the joint test of the two role differences from the pooled specification of Table~\ref{tab:role_difference}, Panel A, with standard errors clustered at the fixed-group level. $^{*}p<0.10$, $^{**}p<0.05$, $^{***}p<0.01$.\par}

\hspace{1cm}

% ================================================================
% M. CITIZENS: OUTCOME OF OWN BRIBE OFFERS
% ================================================================

\section{Citizens: Outcome of Own Bribe Offers}
\label{app:own_offer}

The citizen measures used in the main analysis do not include the outcome of the citizen's own bribe offers, which is the most direct experience of the judge's response to bribery available to a citizen. Table~\ref{tab:own_offer_outcome} adds three measures of this outcome. Of the 295 citizens, 281 make at least one positive offer; on average 11 percent of their positive offers are rejected, and 34 percent of citizens have at least one offer rejected.

None of the three measures is associated with perceived corruption. The coefficient on the share of own offers rejected is $-0.101$ ($p=0.537$), the coefficient on the indicator for at least one rejection is $-0.055$ ($p=0.452$), and the coefficient on the amount rejected, conditional on the other behavioral measures, is $-0.004$ ($p=0.944$). The estimates for the share of own offers rejected are imprecise, and the corresponding confidence interval is wide.

\begingroup
\small
\begin{longtable}{p{6.6cm}ccc}
\caption{\normalsize Citizens: outcome of own bribe offers}
\label{tab:own_offer_outcome}\\
\toprule
& (1) & (2) & (3) \\
& Share rejected & Ever rejected & Amount rejected \\
\midrule
\endfirsthead

\multicolumn{4}{l}{\small\textit{Table \thetable\ continued from previous page}}\\
\toprule
& (1) & (2) & (3) \\
& Share rejected & Ever rejected & Amount rejected \\
\midrule
\endhead

\midrule
\multicolumn{4}{r}{\small\textit{Continued on next page}}\\
\endfoot

\bottomrule
\endlastfoot

\multicolumn{4}{l}{\textit{Panel A. Baseline}} \\
\addlinespace

Share of own positive offers rejected
& -0.1007 &  &  \\
& (0.1624) &  &  \\

At least one own offer rejected
&  & -0.0545 &  \\
&  & (0.0723) &  \\

Own amount rejected
&  &  & -0.0043 \\
&  &  & (0.0609) \\
Bribe offered
&  &  & -0.0096 \\
&  &  & (0.0287) \\

Contribution
&  &  & -0.0027 \\
&  &  & (0.0077) \\

Punishment received
&  &  & 0.0023 \\
&  &  & (0.0114) \\
\midrule

\multicolumn{4}{l}{\textit{Panel B. With background controls}} \\
\addlinespace

Share of own positive offers rejected
& -0.0435 &  &  \\
& (0.1513) &  &  \\

At least one own offer rejected
&  & -0.0117 &  \\
&  & (0.0708) &  \\

Own amount rejected
&  &  & 0.0092 \\
&  &  & (0.0564) \\
Bribe offered
&  &  & -0.0004 \\
&  &  & (0.0265) \\

Contribution
&  &  & -0.0035 \\
&  &  & (0.0076) \\

Punishment received
&  &  & -0.0004 \\
&  &  & (0.0103) \\
\midrule
Experimental-condition controls
& Yes & Yes & Yes \\
Participants
& 281 & 295 & 295 \\
Fixed-group clusters
& 99 & 99 & 99 \\

\end{longtable}
\endgroup

{\footnotesize\noindent\textit{Notes:} OLS estimates using one observation per citizen. The dependent variable is perceived corruption, coded from 1 to 4, with higher values indicating greater perceived corruption. Column (1) uses the share of the citizen's own positive bribe offers that the judge rejects and is estimated on citizens who make at least one positive offer. Column (2) uses an indicator equal to one if at least one of the citizen's positive offers is rejected. Column (3) uses the average amount of the citizen's own offers that is rejected per period, together with the three behavioral measures of the main specification. Behavioral measures are averaged within participant over all periods of the bribery game in which the participant takes part in the interaction. Periods of the Exit condition in which the common project is not implemented because two or more citizens withdraw are treated as missing. Panel B additionally controls for age, gender, association membership, voting status, and blood-donation status. Coefficients on controls are suppressed. Standard errors clustered at the fixed-group level are reported in parentheses, because citizens in the same group face the same judge. $^{*}p<0.10$, $^{**}p<0.05$, $^{***}p<0.01$.\par}

\hspace{1cm}

% ================================================================
% N. EXPERIMENTAL INSTRUCTIONS
% ================================================================

\section{Summary of the Experimental Instructions}
\label{app:instructions}

This section summarizes the instructions for the bribery-game task analyzed in the paper. The original materials used the labels \textit{Participant A} and \textit{Participant B}; these correspond, respectively, to the citizen and judge roles used in the paper. Because most of the instructions were identical across experimental conditions, the common material is reported once and the treatment-specific information and feedback structures are then described separately. The summary is based on the original Italian-language instructions administered to participants. The complete instructions, in the original Italian and in English translation, are reported in the Supplementary Material of \citet{PancottoFedericoRighiHauert2026} (section ``Experimental Instructions''). Instructions for the subsequent experimental tasks are not summarized because those tasks are not part of the analysis in this paper.

\subsection{General Instructions}

Participants were welcomed to VERA Lab and instructed to switch off their phones and not to communicate with other participants or with anyone outside the laboratory during the session. They were informed that they could ask the experimenter questions and use paper and pen for notes or calculations. Decisions could be changed until the ``Next'' button was pressed, after which the selected decision became final. Participants were also informed that all choices were anonymous and that group membership would remain anonymous throughout the study.

Participants received a show-up fee of 3 euros and an additional payment depending on their decisions during the experiment. Experimental earnings were expressed in points, with 100 experimental points corresponding to 3 euros. Participants were informed that they could earn up to 25 euros in the study.

Before the incentivized task began, participants were informed that the first part of the study would last between 12 and 15 periods and that they would complete comprehension questions after reading the instructions. Each group consisted of three Participants A and one Participant B. Roles were assigned randomly at the beginning of the game and remained fixed throughout all periods. Each period consisted of three phases.

\subsection{Common Bribery-Game Instructions}

\subsubsection{Phase 1: Contribution to the Common Project}

At the beginning of each period, each participant received 20 points. Participants A decided how many of their 20 points to allocate to a common project and how many to retain in a private account. Each point not allocated to the common project was automatically credited to the private account.

Each point retained in the private account generated one point of income. The common-project return was determined by multiplying the sum of the three Participants A's contributions by 1.6 and dividing the resulting amount equally among the group members participating in the common project. When all three Participants A participated, the common-project income for each of the four group members was therefore:
\[
\text{Income from common project}
=
\frac{1.6 \times \text{sum of contributions}}{4}.
\]

For example, if the three Participants A contributed 20 points each, total contributions were 60 points and each group member received $(60 \times 1.6)/4=24$ points from the common project. If total contributions were 20 points, each group member received $(20 \times 1.6)/4=8$ points. If nobody contributed, nobody received income from the common project. Decimal values were rounded down when below 0.5 and rounded up when equal to or above 0.5.

When all group members participated, the income of Participant A after Phase 1 was:
\[
20-\text{own contribution}+\frac{1.6 \times \text{sum of contributions}}{4}.
\]

Participant B did not make a decision in Phase 1.

\subsubsection{Phase 2: Transfer to Participant B}

After Phase 1, each Participant A received an additional endowment of 5 points and decided whether and how many of these points, from 0 to 5, to transfer to Participant B. Points not transferred remained part of Participant A's income. Each point transferred cost Participant A one point and, if subsequently accepted by Participant B, increased Participant B's income by one point.

Before the decision of Participant B, the income of Participant A was therefore:
\begin{multline*}
20-\text{own contribution}+\text{income from common project} \\
+5-\text{points transferred to Participant B}.
\end{multline*}

Participant B did not make a decision in Phase 2.

\subsubsection{Phase 3: Acceptance or Rejection and Punishment}

In Phase 3, Participant A did not make any further decision. Participant B was shown separately the contribution to the common project and the transfer offered by each Participant A. For each transfer, Participant B decided whether to accept or reject it. If the transfer was accepted, the corresponding points were credited to Participant B. If it was rejected, the points were returned to the Participant A who had offered them.

The order in which Participants A were displayed to Participant B was randomized. Participants were explicitly informed that a displayed position in one period therefore did not necessarily correspond to the same Participant A in another period.

Participant B also received 15 deduction points. These points could be used to reduce the incomes of Participants A. Each deduction point assigned cost Participant B one point and reduced the targeted Participant A's income by three points. For example, assigning three deduction points reduced Participant A's income by nine points and Participant B's income by three points. Although Participant B had 15 deduction points in total, no more than 10 deduction points could be assigned to any one Participant A. Unused deduction points were credited to Participant B's income for the period.

When all group members participated, the final period income of Participant A was:
\begin{multline*}
20-\text{own contribution}+\text{income from common project} \\
+5-\text{accepted transfer}-3\times\text{deduction points received},
\end{multline*}
where the accepted transfer equals the amount offered if Participant B accepts it and zero if the offer is rejected.

The final period income of Participant B was:
\begin{multline*}
20+\text{income from common project} \\
+\text{sum of accepted transfers}+15-\text{sum of deduction points assigned}.
\end{multline*}

\subsection{Experimental Conditions}

\subsubsection{Full Information (FULL)}

The Full Information condition retained the common three-phase structure described above. In this condition, Participants A received feedback after the contribution stage on the amount generated by the public good and their Phase-1 payoff before proceeding to the transfer decision. At the end of the period, the feedback available within the group included all Participants A's contributions, all transfers offered to Participant B, whether those transfers had been accepted or rejected, and the punishment assigned to Participants A. Thus, Participants A received direct information about bribery transactions involving the other citizens in their group in addition to information about contributions and punishment.

\subsubsection{Control (PINFO)}

The Control condition retained the same contribution, transfer, acceptance\slash rejection, and punishment decisions but restricted the information available to Participants A about bribery involving the other citizens. At the end of each period, Participants A were shown the summary of their own decisions and outcomes together with the contributions of the other Participants A and the deduction points assigned by Participant B. The order in which other Participants A were displayed was randomized across periods. Participants A were not shown the other citizens' transfers to Participant B or whether those transfers were accepted or rejected. Participant B, by virtue of the role, observed the contributions and transfers of all Participants A and made all acceptance\slash rejection and punishment decisions.

\subsubsection{Transparency (PENDO)}

The Transparency condition retained the same basic feedback structure as the Control condition and introduced endogenous access to additional information. Beginning in the second period and before making the Phase-1 contribution decision, each Participant A could pay 3 points to obtain information about Participant B's decisions concerning the other Participants A in the previous period.

The information available for purchase concerned the previous period and included the other Participants A's contributions to the common project, their transfers to Participant B, whether those transfers were accepted or rejected, and the deduction points assigned by Participant B. Participants A decided in each eligible period whether to purchase this information. If purchased, the 3-point information cost was subtracted from the participant's period earnings.

For a Participant A who purchased information, the period-income expression therefore became:
\begin{multline*}
20-\text{information cost}-\text{own contribution}+\text{income from common project} \\
+5-\text{accepted transfer}-3\times\text{deduction points received}.
\end{multline*}

Apart from this information-purchase stage, the contribution, transfer, acceptance\slash rejection, punishment, and end-of-period feedback structure was the same as in the Control condition.

\subsubsection{Exit (PENDO\_EXIT)}

The Exit condition retained the endogenous information-purchase mechanism of the Transparency condition. Beginning in the second period, Participants A could pay 3 points to observe information from the previous period concerning the contributions, transfers, acceptance\slash rejection decisions, and punishment involving the other Participants A.

In addition, Participants A could choose not to participate in the common project in a given period. A Participant A who did not participate did not receive income from the common project and did not take part in the contribution, transfer, or punishment interaction for that period. The income of a Participant A choosing the outside option was:
\[
20+5-\text{information cost},
\]
where the information cost was zero if no information had been purchased.

If exactly one Participant A did not participate, the common project was implemented by the two remaining Participants A and Participant B, and its proceeds were divided among those three participants. In this case Participant B received 10 deduction points rather than 15.

If two or more Participants A did not participate, the common project was not implemented and transfers did not take place. Any transfer that otherwise would have been made was returned to Participant A. Participants A retained their initial 20-point endowment and the additional 5-point endowment, subject to any information cost already incurred, and Participant B retained the initial 20-point endowment.

For a participating Participant A in the Exit condition, period income was:
\begin{multline*}
20-\text{information cost}-\text{own contribution}+\text{income from common project} \\
+5-\text{accepted transfer}-3\times\text{deduction points received}.
\end{multline*}

For Participant B, when the common project was implemented, period income was:
\begin{multline*}
20+\text{income from common project}+\text{sum of accepted transfers} \\
+\text{available deduction points}-\text{sum of deduction points assigned},
\end{multline*}
where the available deduction-point endowment was 15 when all three Participants A participated and 10 when exactly one Participant A did not participate. If the common project was not implemented because two or more Participants A did not participate, Participant B's period income was 20 points.

\subsection{Comprehension Questions and Decision Prompts}

Before the incentivized decisions began, participants completed comprehension questions. The common instructions presented the following example. Participants were asked to imagine that each of the other two Participants A contributed 10 points to the common project and that they, as Participant A, also contributed 10 points. They then answered the following questions:

\begin{enumerate}
\item How many points do you obtain at the end of the period if Participant B does not take away any points from you?
\begin{enumerate}
\item $30\times1.6=48$; $48/4=12$; $20+12=32$.
\item $30\times1.6=48$; $48/4=12$; $20-10+12+5=27$.
\end{enumerate}
\item How many points do you obtain at the end of the period if Participant B reduces your income by 9 points?
\begin{enumerate}
\item $20+12-9=23$.
\item $20-10+12+5-9=18$.
\end{enumerate}
\item How many points do you obtain if you transfer 2 points to Participant B?
\begin{enumerate}
\item $20-10+12+5-2=25$.
\item $20-10+12-2=20$.
\end{enumerate}
\end{enumerate}

The decision screen for Participant A asked how many points, from 0 to 20, the participant wanted to put into the common project. In Phase 2, Participant A was reminded of the additional 5-point endowment and chose a transfer to Participant B from 0 to 5 points. Participant B was then shown each Participant A's contribution and transfer separately and, for each Participant A, chose whether to accept or reject the transfer and how many deduction points, from 0 to 10, to assign.

In the Transparency and Exit conditions, from the second period onward Participants A were additionally asked whether they wanted to purchase information about Participant B's decisions in the previous period at a cost of 3 points. In the Exit condition, Participants A also made the participation decision associated with the outside option before proceeding with the contribution stage.

\bibliographystyle{apalike}
\bibliography{Online_Appendix}